\documentclass[reprint,
superscriptaddress,
 amsmath,amssymb,
 aps,
]{revtex4-2}

\usepackage{graphicx}
\usepackage{dcolumn}
\usepackage{bm}
\usepackage{chemformula}
\usepackage{amsmath,amsthm,amssymb}
\usepackage{gensymb}
\usepackage{hhline}
\usepackage{enumerate}
\usepackage{soul}

\begin{document}

\title{Tracking the local order parameter through the Hubbard exciton decoherence time in the Mott-Hubbard insulator \ch{LaVO3}}

\author{Alessandra Milloch}
\email[]{alessandra.milloch@unicatt.it}
\affiliation{Department of Mathematics and Physics, Università Cattolica del Sacro Cuore, Brescia I-25133, Italy}
\affiliation{ILAMP (Interdisciplinary Laboratories for Advanced
Materials Physics), Università Cattolica del Sacro Cuore, Brescia I-25133, Italy}
\affiliation{Department of Physics and Astronomy, KU Leuven, B-3001 Leuven, Belgium}

\author{Paolo Franceschini}
\affiliation{CNR-INO (National Institute of Optics), via Branze 45, 25123 Brescia, Italy}
\affiliation{Department of Information Engineering, University of Brescia, Brescia I-25123, Italy}

\author{Pablo Villar-Arribi}
\affiliation{International School for Advanced Studies (SISSA), Via Bonomea 265, Trieste 34136, Italy}

\author{Sandeep Kumar Chaluvadi}
\affiliation{CNR-IOM (Istituto Officina dei Materiali) S.S. 14, km 163.5, I-34149 Trieste, Italy}

\author{Pasquale Orgiani}
\affiliation{CNR-IOM (Istituto Officina dei Materiali) S.S. 14, km 163.5, I-34149 Trieste, Italy}

\author{Giancarlo Panaccione}
\affiliation{CNR-IOM (Istituto Officina dei Materiali) S.S. 14, km 163.5, I-34149 Trieste, Italy}

\author{Giorgio Rossi}
\affiliation{CNR-IOM (Istituto Officina dei Materiali) S.S. 14, km 163.5, I-34149 Trieste, Italy}
\affiliation{Dipartimento di Fisica, Università degli Studi di Milano, Via Celoria 16, 20133 Milano, Italy}

\author{Yang Liu}
\affiliation{Laboratory of Atomic and Solid State Physics, Cornell University, Ithaca, NY 14853, USA}

\author{Darrell G. Schlom}
\affiliation{Department of Materials Science and Engineering, Cornell University, Ithaca, New York 14853, USA}
\affiliation{Kavli Institute at Cornell for Nanoscale Science, Ithaca, New York 14853, USA}
\affiliation{Leibniz-Institut für Kristallzüchtung, Max-Born-Str. 2, 12489 Berlin, Germany}

\author{Kyle M. Shen}
\affiliation{Laboratory of Atomic and Solid State Physics, Cornell University, Ithaca, NY 14853, USA}
\affiliation{Kavli Institute at Cornell for Nanoscale Science, Ithaca, New York 14853, USA}

\author{Massimo Capone}
\affiliation{International School for Advanced Studies (SISSA), Via Bonomea 265, Trieste 34136, Italy}
\affiliation{CNR-IOM (Istituto Officina dei Materiali), Via Bonomea 265, Trieste 34136, Italy}

\author{Claudio Giannetti}
\email[]{claudio.giannetti@unicatt.it}
\affiliation{Department of Mathematics and Physics, Università Cattolica del Sacro Cuore, Brescia I-25133, Italy}
\affiliation{ILAMP (Interdisciplinary Laboratories for Advanced
Materials Physics), Università Cattolica del Sacro Cuore, Brescia I-25133, Italy}
\affiliation{CNR-INO (National Institute of Optics), via Branze 45, 25123 Brescia, Italy}

\begin{abstract}
The prototypical Mott-Hubbard insulator \ch{LaVO_3} undergoes a structural phase transition accompanied by the onset of spin and orbital ordering below 140 K. By combining ultrafast optical pump-probe spectroscopy and two-dimensional electronic spectroscopy, we investigate the interplay between fluctuations of the local spin and orbital order parameter and the lifetime of high-energy electron-hole excitations. Specifically, we demonstrate that the pump-induced perturbation of the order parameter leads to a change of the Hubbard exciton decoherence time and, consequently, of its homogeneous linewidth. Dynamical mean-field theory calculations confirm that the exciton scattering rate is crucially affected by the degree of order of the spin and orbital lattices in \ch{LaVO3}. Our results demonstrate that multi-dimensional ultrafast optical spectroscopy can be used to track the dynamics of the order parameter, thus opening new routes in the study of correlated quantum materials characterized by intertwined orders.

\end{abstract}
\maketitle


\section{Introduction}




Strongly interacting materials are intrinsically prone to the development of local correlations in the charge/spin/orbital channels, which eventually drive complex symmetry-breaking phase transitions \cite{imada1998metal,tokura2003PhysToday,dagotto2005complexity, dagotto2008strongly,basov2011electrodynamics}. 
In the attempt to track the dynamics of order parameters, recent efforts have been devoted to the development of time-resolved probes sensitive to the specific degrees of freedom that are relevant for the phase transition. Time-resolved X-ray diffraction, resonant X-ray scattering and electron diffraction have been used to track the lattice order parameter dynamics, as well as charge and orbital degrees of freedom in systems with long-range order \cite{beaud2014time, vogelgesang2018phase,zong2019dynamical,zhu2018unconventional,mitrano2019ultrafast,trigo2021ultrafast,huber2014coherent,maklar2021nonequilibrium}. Second harmonic generation and reflectivity anisotropy experiments have been introduced to measure the symmetry of the order parameter and its temporal dynamics \cite{perez2022multi, tzschaschel2019tracking}. Order parameters can also be indirectly probed via time- and angle-resolved photoemission spectroscopy (tr-ARPES) for systems in which the emerging order is coupled to the electronic band structure and leads to the opening of a gap \cite{zong2019evidence,huber2022revealing,maklar2021nonequilibrium}. On top of these approaches, great attention has been given to all-optical spectroscopy experiments, which probe the electronic properties and their dynamics during the photoinduced melting of order parameters. In many cases, the electronic dynamics measured by transient optical reflectivity experiments have been reported to be representative of the time evolution of the order parameters \cite{yusupov2010coherent,kim2024emergent,ning2023coherent,lovinger2020,zhu2018unconventional}. Nonetheless, while optical probes represent a simple and highly attractive approach, the link between the reflectivity dynamics and the order parameter remains highly indirect.

\begin{figure*}[]
\includegraphics[width=15cm]{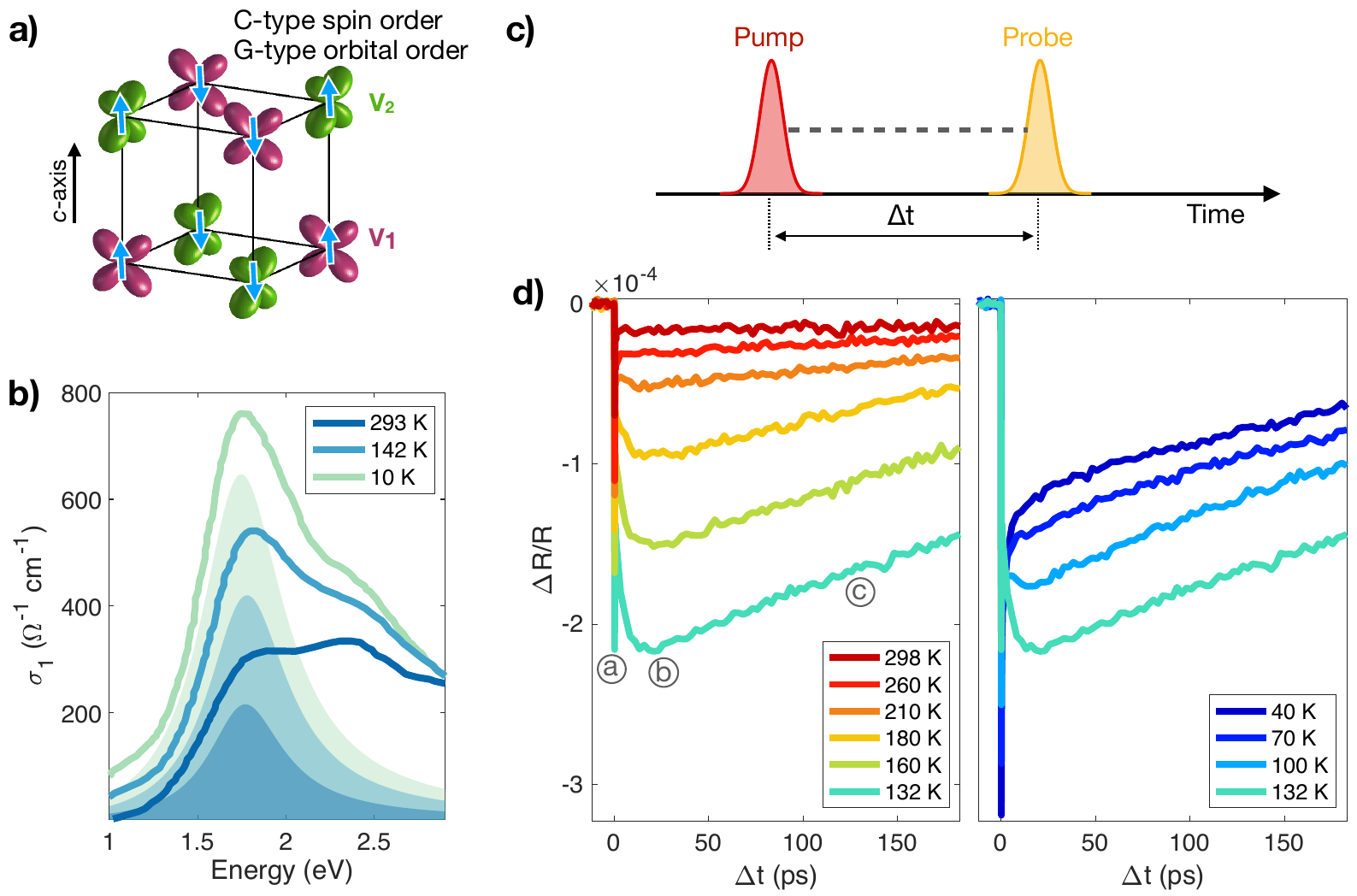}
\caption{a)  Spin and orbital ordered structure of \ch{LaVO_3} for $T < T_c$. b) Equilibrium optical conductivity (real part $\sigma_1$) of \ch{LaVO_3} for light polarization parallel to the $c$-axis at three different temperatures across $T_c$ (from Ref. \citenum{miyasaka2002}), fitted using a multi-peak Drude-Lorentz model. The lowest energy component, highlighted by the filled area in the graph, is associated with a Hubbard exciton resonance. c) Sketch of a pump-probe experiment: the pump pulse generates an out-of-equilibrium state, whose time-evolution is detected by the probe (delayed by $\Delta t$) that measures the pump-induced changes in the reflectivity of the sample. d) Optical pump-probe measurements in reflection geometry at different temperatures, above (left panel) and below (right panel) $T_c$. The measurements have been performed with 1.65~eV photon energy pump, 1.77~eV photon energy probe, and 0.1~mJ/cm$^2$ excitation fluence.}
\label{fig1: LaVO3 intro}
\end{figure*}

In this work, we combine broadband pump-probe spectroscopy, two-dimensional electronic spectroscopy (2DES), and dynamical mean-field theory (DMFT) to demonstrate that the intrinsic lifetime of well-defined electronic excitations, coupled to degrees of freedom undergoing a symmetry-breaking phase transition, provides a direct mapping of the order parameter. To achieve this goal, we consider the prototypical case of the Mott-Hubbard insulator \ch{LaVO3}, a transition-metal oxide whose optical properties are characterized by an excitonic resonance, ascribed to a Hubbard exciton (HE) \cite{kim2018signatures}. 
By using ultrafast broadband and multidimensional spectroscopies, we demonstrate that the HE decoherence time is affected by the onset of long-range spin and orbital ordering. Ultrafast optical spectroscopy can thus be used to investigate the dynamics of the order parameter over a broad range of temperatures across the spin and orbital ordering first-order phase transition at $140$~K. The dynamics is empirically well captured by a Ginzburg-Landau-inspired model describing an order parameter coupled to electronic excitations. DMFT calculations are then used to corroborate the link between the orbital and spin correlations and the HE scattering rate in a multi-band model. 


The work is organized as follows. After introducing the spin and orbital ordered phase in \ch{LaVO3} (Sec. II), we discuss in Sec. III pump-probe experiments where a broadband probe that covers the HE spectral region (1.3-2.0~eV) is used to spectrally resolve the transient reflectivity. This broadband approach expands the information obtainable from single-color experiments \cite{lovinger2020} and enables disentangling spectral weight variations from changes in the width or energy of the excitonic resonance. Our data reveal that, in addition to a decrease in spectral weight, the HE undergoes a linewidth broadening on a timescale of tens of picoseconds. We observe that the HE linewidth follows the dynamics of the order parameter, as predicted by the Ginzburg-Landau description of the spin- and orbital-ordering phase transition, discussed in Sec. IV, where the order parameter couples to the photo-induced 
 electronic excitations. 
To gain further insight into the nature of the revealed HE linewidth broadening, we employ 2DES, an ultrafast spectroscopy technique that provides access to the intrinsic homogeneous linewidth, disentangling it from inhomogeneous contributions to spectral broadening \cite{jonas2003two,cho2008coherent,mukamel2000multidimensional}. 2DES experiments on \ch{LaVO3} (Sec. V) reveal that the photo-induced broadening of the HE linewidth results from a broadening of its homogeneous component, providing direct evidence of the decrease of  HE decoherence time caused by the pump-induced perturbations. In Sec. VI, we finally rationalize the mapping of the order parameter onto the HE intrinsic lifetime. We present a three-band model DMFT calculation for \ch{LaVO3} showing that (i) the scattering rate increases upon suppression of the long-range spin and orbital order, in agreement with ultrafast optical spectroscopy, and (ii) it displays a temperature-dependent behavior, upon approaching the phase transition critical temperature, consistent with the Ginzburg-Landau order parameter. These results therefore demonstrate that the HE lifetime is a good representative of the order parameter, thus enabling tracking of its evolution.  

\section{Spin and orbital order in L\lowercase{a}VO$_3$}

The class of vanadium oxides \ch{RVO_3} (R = rare-earth or Y) represents a group of prototypical Mott insulators considered ideal systems for investigating the interplay between electronic excitations and spin, orbital, and lattice degrees of freedom \cite{kim2018signatures,fang2004quantum,de2007orbital,khaliullin2005orbital,horsch2003dimerization,miyasaka2000critical}. In these compounds, the \ch{t_{2g}^2} electronic configuration of the trivalent vanadium cation results in exotic spin and orbital ordering phases \cite{ulrich2003magnetic,tung2008spin,sawada1996orbital}. Spin and orbital degrees of freedom also influence electronic excitations across the Mott-Hubbard gap favoring the formation of bound states between an empty site (hole) and a doubly occupied one (doublon), known as Hubbard excitons (HEs) \cite{essler2001excitons,matsueda2005excitonic,gossling2008mott,matiks2009exciton,novelli2012ultrafast,reul2012probing}. Stabilization of the HE is typically governed by the gain in kinetic energy for the bound state in the spin/orbital ordered environment \cite{clarke1993particle,wrobel2002excitons}. 

In this work, we focus specifically on \ch{LaVO_3} \cite{kim2018signatures,lovinger2020}. \ch{LaVO_3} has a perovskite-type structure and it undergoes a first-order structural phase transition from orthorombic (space group $Pnma$) to monoclinic lattice (space group $P2_1/a$) at the critical temperature $T_c$~=~140~K \cite{bordet1993}. Below $T_c$, the lattice is also subject to Jahn-Teller elongation, in addition to the \ch{GdFeO_3}-type distortion already present for $T > T_c$ \cite{bordet1993,deraychaudhury2007}. These structural variations affect in turn the spin and the orbital degrees of freedom \cite{horsch2008}: around the same critical temperature $T_c$, the system becomes both magnetically and orbitally ordered, with C-type and G-type order respectively. This means that, as sketched in Fig. \ref{fig1: LaVO3 intro}a, there is a ferromagnetic stacking (along the $c$-axis) of antiferromagnetic planes, along with an alternated occupation of $d_{zx}$ and $d_{yz}$ orbitals of \ch{V}-sites along all crystallographic axis (whereas $d_{xy}$ is occupied at all \ch{V}-sites) \cite{deraychaudhury2007, zhou2008}. 
The lowest-energy electronic excitation across the Mott-Hubbard gap, which corresponds to the transition of a $d_{yz}$ or $d_{zx}$ electron on a V-site to an adjacent vanadium site along the $c$-axis, can result in a HE bound state. In the optical conductivity spectrum of \ch{LaVO_3} (see Fig. \ref{fig1: LaVO3 intro}b), the HE resonance gives rise to a peak at 1.8~eV, which is enhanced below the critical temperature as the HE is stabilized by the spin and orbital order of the system \cite{miyasaka2002}.

\section{Pump-probe spectroscopy}
\label{sec:pump-probe lvo} 

\begin{figure*}
\includegraphics[width=13cm]{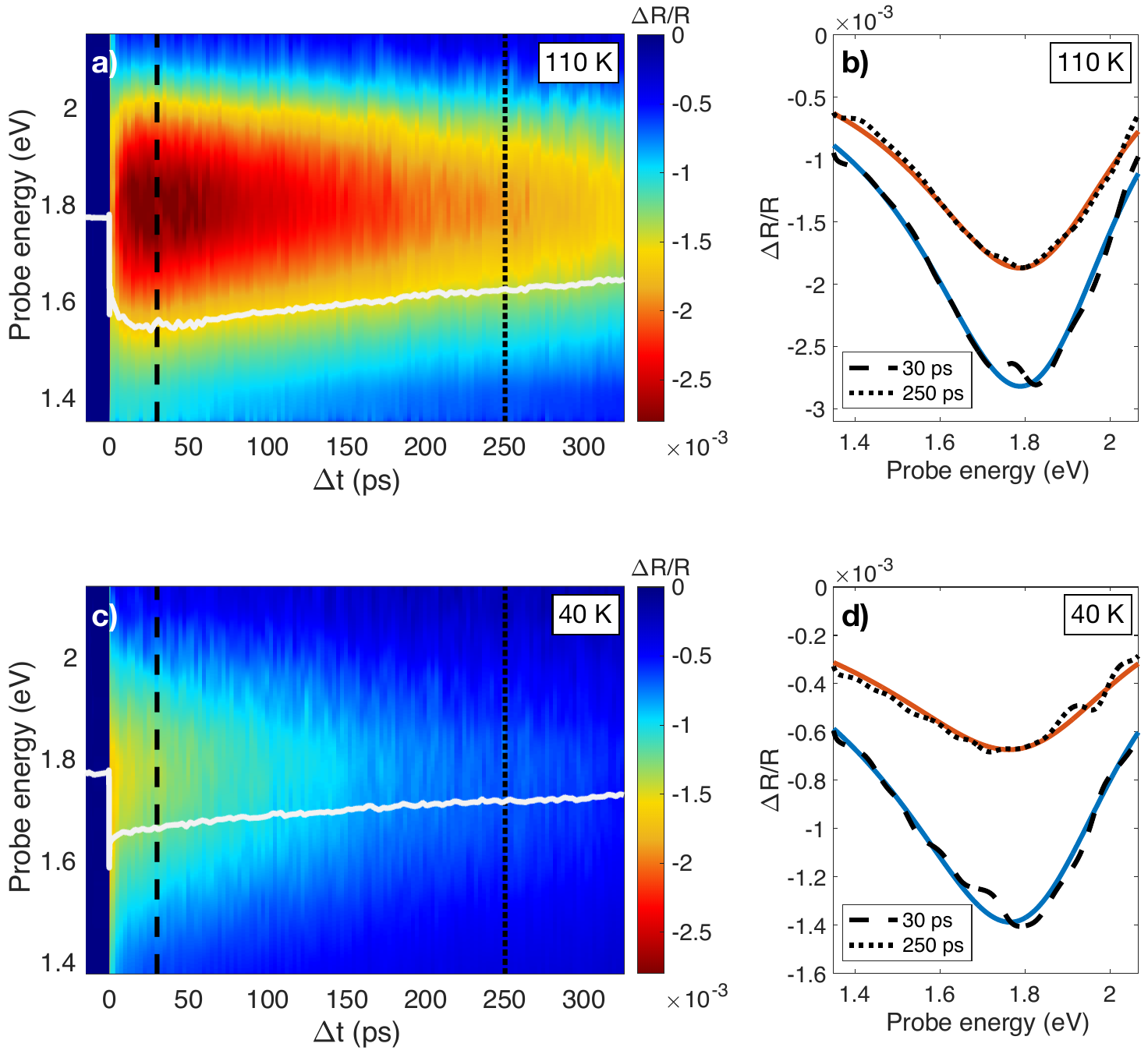}
\caption{Broadband pump-probe measurements performed with 1.4~eV photon energy pump and supercontinuum probe in reflection geometry. The transient reflectivity spectra are plotted as a function of pump-probe time delay (horizontal axis) and probe photon energy (vertical axis) in a) and c), for sample temperatures of 110~K and 40~K, respectively. The pump incident fluence is 1~mJ/cm$^2$. The white lines plotted on top represent the time evolution of the signal at 1.77~eV and display the same behavior described in Fig. \ref{fig1: LaVO3 intro}d. Panels b) and d) report the transient reflectivity spectra at two selected delays, highlighted in a) and c) by black dashed (30~ps) and dotted (250~ps) lines. Red and blue solid lines are differential fits of the $\Delta R/R$ spectra, performed as described in Appendix C. }
\label{fig3: dRR_maps}
\end{figure*}

Ultrafast optical pump-probe measurements (see Fig. \ref{fig1: LaVO3 intro}c) can be used to investigate the out-of-equilibrium behavior of the electronic excitations in \ch{LaVO_3} \cite{tomimoto2003ultrafast,lovinger2020} via monitoring the photo-induced variation of the materials optical properties. 
Fig. \ref{fig1: LaVO3 intro}d shows the single-color transient reflectivity signal ($\Delta R/R$) measured at 1.77~eV probe photon energy for various temperatures below and above the transition temperature (right and left panel, respectively).
The out-of-equilibrium dynamics displays three main components: a nearly instantaneous electronic response, decaying within 500~fs ((a) in Fig. \ref{fig1: LaVO3 intro}d), an intermediate component raising on $\sim$~20~ps timescale ((b) in Fig. \ref{fig1: LaVO3 intro}d) and a slow recovery taking place over hundreds of picoseconds ((c) in Fig. \ref{fig1: LaVO3 intro}d). The (a) component is typically assigned to photocarrier injection by intersite vanadium $d-d$ excitations, relaxing over $\sim$~500~fs via electron-electron and electron-phonon scattering \cite{lovinger2020}, whereas the slower signal is related to the HE resonance. Fig. \ref{fig1: LaVO3 intro}d shows that, in agreement with previous experimental reports \cite{lovinger2020}, the picosecond component (b) strongly depends on temperature: it is enhanced when the system is close to the critical temperature $T_c$, while it disappears both at room temperature (dark red line) and at low temperatures (dark blue line). This component has been ascribed to spectral weight transfer away from the HE after pump excitation and back to it during the recovery dynamics (c) \cite{lovinger2020}. It has been suggested that such spectral weight changes are determined by spin and orbital fluctuations that become larger in proximity of $T_c$ or due to pump-induced spin/orbital disorder \cite{lovinger2020}.  At low temperatures, the HE pump-probe signal is instead suppressed because spin and orbital degrees of freedom freeze, strong order is established and spin/orbital fluctuations are reduced \cite{lovinger2020}. Similarly, the transient reflectivity signal depends on the excitation fluence (see Fig. S1 in Supplemental Material), with the (b) component being enhanced as the excitation intensity is increased. This suggests that high excitation fluence enhances the disruption of the ordered background, with larger intensities required to perturb the spin and orbital order as the temperature is further lowered below $T_c$.

Here, we extend the investigation of the HE dynamics on the picosecond timescale by performing broadband pump-probe spectroscopy on a \ch{LaVO_3} thin film. We employ a 1.4~eV photon energy pump with 50 fs time duration, and a supercontinuum probe in the 1.3-2.0~eV spectral range (see Appendix B for experimental details). 
The transient reflectivity data after correction of the chirp of the probe pulse are reported as colormaps in Fig. \ref{fig3: dRR_maps} as a function of probe photon energy and pump-probe time delay, $\Delta t$, for two different temperatures below $T_c$: 110~K and 40~K (panels a and c, respectively).
Consistently with single-color pump-probe measurements (Fig. \ref{fig1: LaVO3 intro}d) and previous experimental reports \cite{lovinger2020}, we can observe an ultrafast electronic response, followed by the slower dynamics taking place on a timescale spanning 10-100~ps. 
This long-lived signal displays the expected temperature-dependent behavior characterized by the intermediate component ((b) in Fig. \ref{fig1: LaVO3 intro}d, building up over a few tens of picoseconds) that emerges in proximity of $T_c$ (see white solid lines in Figs. \ref{fig3: dRR_maps}a and \ref{fig3: dRR_maps}c). The use of a supercontimuum probe reveals that this slow and negative component of $\Delta R/R$ is a broad response centered at the HE energy. The spectral response at selected pump-probe time delays, namely  30 and 250~ps, obtained from vertical line cuts of Figs. \ref{fig3: dRR_maps}a and \ref{fig3: dRR_maps}c, is plotted in Figs. \ref{fig3: dRR_maps}b and \ref{fig3: dRR_maps}d for the two temperatures, respectively. 

The broadband transient reflectivity data are analyzed by performing a differential fit (see red and blue solid lines in Figs. \ref{fig3: dRR_maps}b and \ref{fig3: dRR_maps}d). For each time delay $\Delta t \geq $~500~fs, a differential reflectivity $(R^{neq}-R^{eq})/R^{eq}$ is fitted to the measured spectral response at fixed $\Delta t$, where $R^{eq}$ and $R^{neq}$ are the equilibrium and out-of-equilibrium reflectivities of the sample, computed from a multi-peak Drude-Lorentz parametrization of the \ch{LaVO3} dielectric function. The free parameters of the fit procedure are the HE plasma frequency, $\omega_{p,HE}$
, and the HE linewidth, $\Gamma_{HE}$. A detailed discussion of the fitting analysis is reported in Appendix C. 
The analysis reveals that the ultrafast reflectivity variation is not simply related to the exciton spectral weight transfer; the ultrafast photo-excitation has, in fact, two concurrent effects on the HE resonance: 
\begin{enumerate}[(i)]
\item a decrease of excitonic spectral weight; 
\item a broadening of the HE peak linewidth $\Gamma_{HE}$.
\end{enumerate}


 


 \begin{figure*}[t]
\includegraphics[width=16cm]{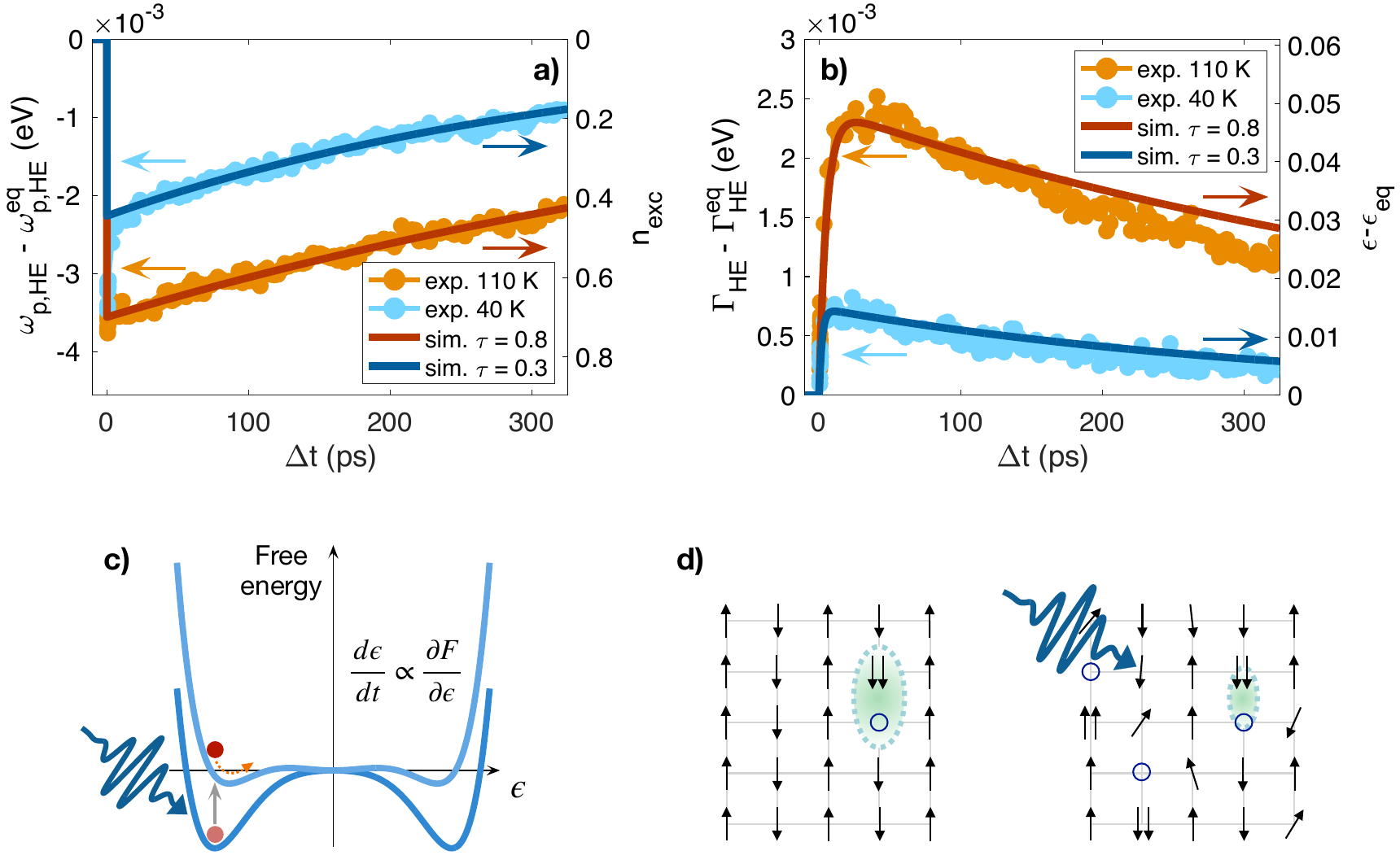}
\caption{a,b) Time evolution of the parameters describing the out-of-equilibrium state of the HE component, retrieved from the fitting analysis of the $\Delta R/R$ data in Fig.~\ref{fig3: dRR_maps}.
In panel a), the variation of Hubbard exciton's plasma frequency, $\omega_{p,HE}-\omega_{p,HE}^{eq}$ ($\omega_{p,HE}^{eq}$ being the equilibrium value), is plotted as light blue and orange dots (left $y$-axis) for the two measured temperatures (40~K and 110~K, respectively), and is compared to the decay of the number of electronic excitations $n_{exc}$ (red and blue solid lines, right $y-$axis). Similarly, in panel b), the variation of Hubbard exciton's width $\Gamma_{HE}-\Gamma_{HE}^{eq}$ (light blue and orange dots, left $y$-axis), $\Gamma_{HE}^{eq}$ being the equilibrium linewidth, is compared to the dynamics of the order parameter $\epsilon$ (red and blue solid lines, right $y-$axis), obtained from numerical integration of Eq. \ref{eq: eps time evo} with $g = 0.6$, $ \alpha = 5$, $\gamma = 0.01$~ps$^{-1}$. 
c) Sketch of the free energy for a first-order phase transition as a function of the order parameter $\epsilon$ and its pump-induced perturbation. The photoexcited electronic population couples to the order parameter, whose dynamics is governed by the free energy potential. d) Cartoon of the Hubbard exciton in the spin and orbital ordered background. Disruption of the ordered background due to pump excitation leads to a reduction of the intrinsic lifetime (decoherence time). }
\label{fig4: param timeevo simul}
\end{figure*}

Figs. \ref{fig4: param timeevo simul}a and \ref{fig4: param timeevo simul}b show the retrieved temporal dynamics of the parameters describing the out-of-equilibrium state of HE. 
In Fig. \ref{fig4: param timeevo simul}a, the orange and light blue circles show the dynamics of the plasma frequency $\omega_{p,HE}$ of the HE Drude-Lorentz oscillator extracted for $T$~=~110~K and 40~K, respectively. 
The ultrafast reduction in $\omega_{p,HE}$, following the impulsive pump excitation, represents the decrease of excitonic spectral weight. At high excitation intensities, this effect originates from the pump-induced destabilization of the ordered background, which undermines the stability of bound excitonic pairs and leads to the population of unbound electronic excitations (holons and doublons). The transient decrease of the HE spectral weight can therefore be regarded as a direct representation of the number of {unbound} electronic excitations, $n_{exc}$, generated by the pump excitation. At low temperatures, where the HE is more robust (see Fig. \ref{fig1: LaVO3 intro}b), the formation of unbound holons and doublons is hindered, resulting in a smaller pump-induced change in the spectral weight compared to temperatures closer to $T_c$.
Such electronic population returns to the equilibrium state within hundreds picoseconds (see Fig. \ref{fig4: param timeevo simul}a), so it can overall be described as an exponential decay 
\begin{equation}
    n_{exc} = n_{exc,0} \theta(t) e^{-t/\tau_{exc}}
    \label{eq: num exc}
\end{equation}
where $\theta(t)$ is the Heaviside step function and $\tau_{exc}$ is the characteristic relaxation time. An exponential fit of $\omega_{p,HE}-\omega_{p,HE}^{eq}$ (where $\omega_{p,HE}^{eq}$ represents the equilibrium value of the HE plasma frequency), returns $\tau_{exc}$~=~350~ps at 40~K and 650~ps at 110~K.

On the other hand, the temporal dynamics of the linewidth of the HE peak displays a significantly different behavior (Fig. \ref{fig4: param timeevo simul}b). Indeed, compared to the $\omega_{p,HE}$ temporal profile, the broadening of $\Gamma_{HE}$ is delayed in time, reaching a maximum after $\sim$~20~ps at 40~K and $\sim$~40~ps at 110~K. The following relaxation dynamics instead occurs on a timescale similar to the one of $\omega_{p,HE}$.

\section{Ginzburg-Landau description}
\label{sec: gl}
To assess the origin of the temperature-dependent behavior of the HE dynamics, we model our system within the framework of Ginzburg-Landau theory. As the phase transition in \ch{LaVO3} has a first-order nature \cite{tung2008spin,ren2003orbital,lovinger2020}, we write the free energy as 
  \begin{equation}
     f(\epsilon) = \alpha(\tau-1)\epsilon^2+\epsilon^2(\epsilon^2-1)^2,
\label{eq: free energy eq}
 \end{equation}
 where $\epsilon$ is the order parameter, $\alpha$ is a numerical constant and $\tau=T/T_{c}$ is the normalized temperature. The first-order nature of the process is guaranteed by the negative coefficient of the fourth-order term $\epsilon^4$. In order to account for the coupling of the electronic excitations ($n_{exc}$) to the spin and orbital long-range order, we include in the free energy an additional coupling term $g\epsilon^2 n_{exc}$. Overall, the free energy reads
 \begin{equation}
   f(\epsilon,n_{exc}) = \alpha (\tau-1)\epsilon^2+\epsilon^2(\epsilon^2-1)^2 + g\epsilon^2 n_{exc}
\label{eq: free energy perturbed}
 \end{equation}
with $n_{exc}$ given by Eq. \ref{eq: num exc} and $g$ coupling constant between the number of photoinduced electronic excitations and the order parameter. Fig. \ref{fig4: param timeevo simul}c shows a sketch of the free energy curve at equilibrium (dark blue solid line, Eq. \ref{eq: free energy eq}) and its instantaneous variation (light blue solid line, Eq. \ref{eq: free energy perturbed}) due to coupling with the electronic excitations generated by the light pulse excitation. This light-induced change of the free energy functional results in a perturbation of the order parameter $\epsilon$ because, upon excitation of $n_{exc}$, the free energy minimum shifts from the equilibrium value, $\epsilon_{eq}$, to a different value of $\epsilon$. 

The dynamics of the order parameters $\epsilon$ is then determined by the kinetic equation \cite{giannetti2016}: 
\begin{equation}
    \frac{d\epsilon}{dt} = -\gamma \frac{\partial f}{\partial \epsilon},
    \label{eq: eps time evo}
\end{equation}
with $\gamma$ time constant. No source term appears in the kinetic equation because of no direct coupling between the light and the order parameter of the system, which includes orbital and magnetic degrees of freedom. The coupling $g$ to the electronic excitation is what determines the perturbation of $\epsilon$ after pulse excitation. Because of the free energy flattening around the minimum as the temperature approaches the critical one, the variation of the order parameter displays, for a fixed excitation density $n_{exc}$, a slower build-up dynamics and a larger amplitude when the temperature increases towards $T_c$ (see Fig. S2a-b in Supplemental Material, reporting an example of order parameter dynamics, obtained upon integrating Eq. \ref{eq: eps time evo}, for several normalized temperatures $\tau$).

We now compare the dynamics of the parameters extracted from broadband pump-probe measurements with the predictions of the Ginzburg-Landau model introduced in Eq. \ref{eq: free energy perturbed} and \ref{eq: eps time evo}. We suppose that $\omega_{p,HE}$ maps $n_{exc}$ (see Eq. \ref{eq: num exc}), and $\Gamma_{HE}$ maps the order parameter $\epsilon$. We underline that, in doing so, we make the two following assumptions: 
\begin{enumerate}[(1)]
\item the variation of the HE linewidth measured in pump-probe spectroscopy corresponds to the variation of the intrinsic lifetime;
\item the Hubbard exciton intrinsic lifetime can be considered as representative of the order parameter of the phase transition.
\end{enumerate}
The justification of assumptions (1) and (2) will be discussed in Sec. \ref{sec:2d lvo} and \ref{sec: dmft}, respectively, where multi-dimensional spectroscopy data and DMFT calculations for \ch{LaVO3} will be considered. Under these assumptions, we compare, in Fig. \ref{fig4: param timeevo simul}b, the experimentally estimated $\Gamma_{HE}$ dynamics (see Sec. \ref{sec:pump-probe lvo} and Appendix C) to the time evolution of $\epsilon$. Red and blue solid lines (right axis) are obtained starting from Eq. \ref{eq: free energy perturbed} and \ref{eq: eps time evo}, using a possible set of parameters that allows to match the experimental trend. In particular, the normalized temperature is set to $\tau = $ 0.3 and 0.8 for blue and red plots, respectively, according to the experimental temperature at which the experiment is performed. The time-dependent term $n_{exc}(t)$ is given by the red and blue solid lines in Fig. \ref{fig4: param timeevo simul}a (right axis), having amplitude that matches the experimental trends of $\omega_{p,HE} - \omega_{p,HE}^{eq}$ for the two temperatures (we chose $n_{exc,0} = 0.45$ and 0.7 for $T = 40$~K and 110~K, respectively). The value of the parameters $\gamma$, $\alpha$ and $g$ is then set in order to match the experimental trends of $\Gamma_{HE}$. The parameters $\alpha$ and $g$ appearing in the free energy profile determine the amplitude of $\epsilon-\epsilon_{eq}$, which depends linearly also on $n_{exc,0}$; when $n_{exc,0}$ is fixed as discussed above, setting $\alpha = 5$ and $g = 0.6$ allows to match, simultaneously for both temperatures, the amplitude of the order parameter dynamics and the linewidth variation. Lastly, $\gamma$ determines the build-up time of the dynamics of $\epsilon-\epsilon_{eq}$, with a time constant that depends also on the normalized temperature $\tau$ and on the free energy coefficient $\alpha$. With $\tau$ being fixed to the experimental values and $\alpha$ = 5, we find that the value $\gamma = 0.01$~ps$^{-1}$ allows to reproduce the experimental dynamics at both temperatures.
For the parameters chosen as described above, the free energy $f$ at equilibrium (Eq. \ref{eq: free energy eq}) and right after excitation (Eq. \ref{eq: free energy perturbed}) are plotted in Fig. S2c Supplemental Material.  In this way, the time dependence of the order parameter $\epsilon$, extracted from the kinetic equation, reproduces all the main features observed in the experimental dynamics of the linewidth $\Gamma_{HE}$, as listed below: 
\begin{enumerate}[(i)]
\item The response is delayed due to a finite build-up time of few tens of ps; the subsequent relaxation decay is governed by $\tau_{exc}$. 
\item The perturbation of the order parameter is enhanced when the temperature of the system is closer to $T_c$, as observed from $\epsilon - \epsilon_{eq}$ being a factor 3.3 larger at 110~K compared to 40~K, despite $n_{exc,0}$ being only a factor 1.5 larger at the higher temperature. 
\item The dynamics slows down as the temperature approaches $T_c$, as observed from the longer build-up time that characterizes both $\Gamma_{HE}-\Gamma_{HE}^{eq}$ and $\epsilon-\epsilon_{eq}$ at 110~K as compared to 40~K. 
\item The response scales nonlinearly with $n_{exc,0}$, with an enhanced perturbation of the order parameter at higher excitation intensities (see Supplemental Material Fig. S3), in agreement with the measured fluence-dependent transient reflectivity (see Supplemental Material Fig. S1), whose picosecond component qualitatively tracks the HE linewidth dynamics.
\end{enumerate}

The agreement between the free energy model and the experimental results suggests that the variations in the linewidth $\Gamma_{HE}$ are ascribable to a coupling between the light-induced electron excitations and the order parameter. An improved description of the long timescale (hundreds of picoseconds) recovery dynamics can be obtained by including a direct decay term for the order parameter, consistent with a transient, laser-induced increase of the system temperature, as detailed in Supplemental Material Sec. S3. Regardless of the specifics of modelling, the emerging scenario is that ultrafast excitation of charges across the Mott-Hubbard gap causes a local perturbation of the order parameter, corresponding to an increase of spin/orbital fluctuations, as sketched in Fig. \ref{fig4: param timeevo simul}d, and a broadening of the Hubbard exciton linewidth.





\section{2D electronic spectroscopy}
\label{sec:2d lvo}

\begin{figure*}
\includegraphics[width=17.7cm]{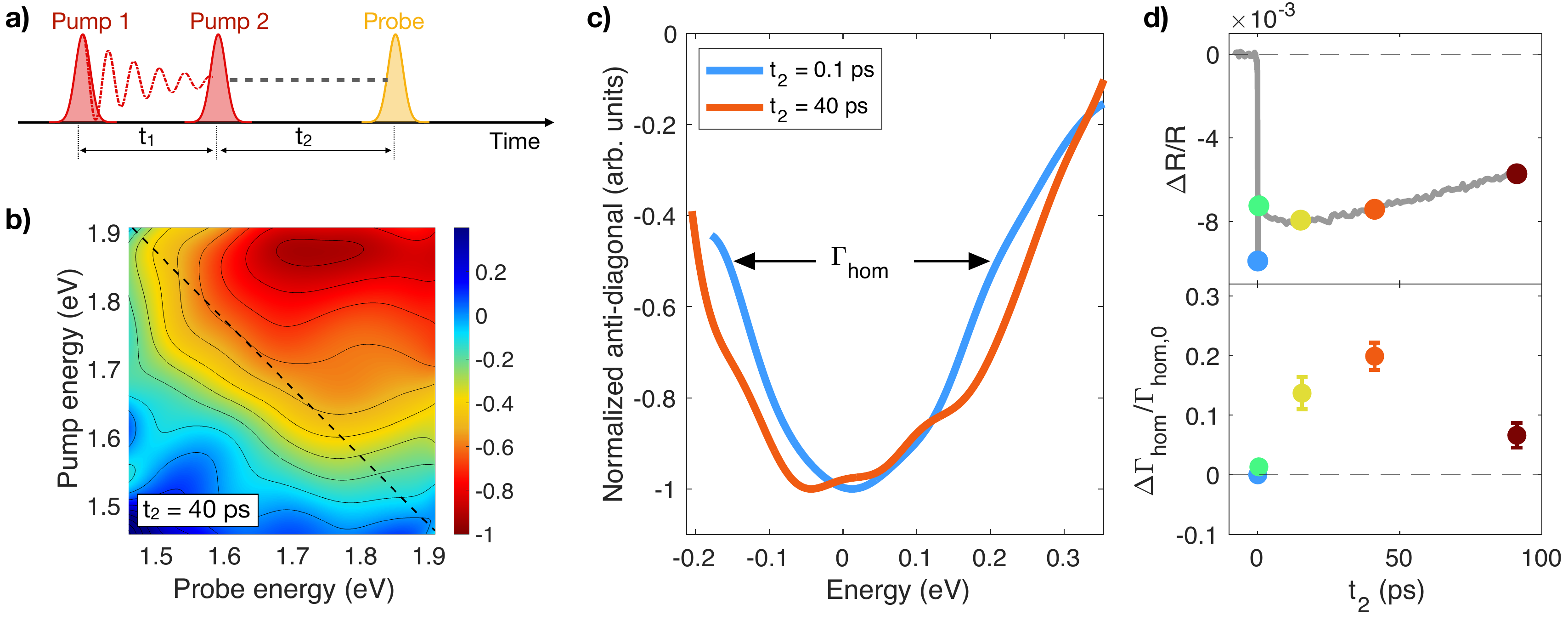}
\caption{a) Sketch of a multi-dimensional spectroscopy experiment, employing two phase-coherent pump pulses, delayed by a variable time delay $t_1$, and a probe pulse delayed by $t_2$. b) 2DES measurement performed at $T = 140$~K, $t_2$~=~40~ps and excitation fluence 1.4~mJ/cm$^2$. The 2D spectrum (arb. units) is normalized over both the probe and pump spectra. c) Anti-diagonal profiles of 2D spectra at two different time delays $t_2$ (red and blue lines); they are obtained from a line-cut along the direction indicated by the black dashed line in b) and are integrated over 25~meV width. The plotted values are normalized in intensity for comparison purposes. d) The top panel reports the pump-probe dynamics (grey line) measured in the same experimental configuration of the 2DES data in b) and c), with broadband (1.45-1.9~eV) and degenerate pump and probe beams at $T = 140$~K and 1.4~mJ/cm$^2$ excitation fluence. The colored dots indicate the $t_2$ delays where 2D spectra are collected. In the bottom panel, the relative variation of the anti-diagonal linewidth ($\Gamma_{hom}$, FWHM) extracted from 2D spectra is plotted as a function of $t_2$. It is estimated as $\Delta \Gamma_{hom} / \Gamma_{hom,0} = [\Gamma_{hom}(t_2)-\Gamma_{hom}(t_2 = 100 ~\text{fs})]/\Gamma_{hom}(t_2 = 100 ~\text{fs})$, where the value obtained at the shortest time delay, $t_2$~=~100~fs, is used as reference ($\Gamma_{hom,0} = \Gamma_{hom}(t_2 = 100 ~\text{fs})$).}
\label{fig5: 2D spectroscpy}
\end{figure*}

The linewidth variations obtained so far from pump-probe data are not directly related to the exciton lifetime because the pump-probe spectral response can be dominated by inhomogeneous broadening. As opposed to pump-probe spectroscopy, 2DES can instead directly access the exciton intrinsic lifetime (decoherence time). We therefore employ 2DES, with the aim of providing insight on the HE decoherence time.

2DES is an ultrafast spectroscopy \cite{biswas2022coherent,collini20212d} technique that extends conventional pump-probe spectroscopy by using two phase-coherent pump pulses, separated by a variable time $t_1$ (coherence time), and a third ultrashort pulse, the probe, delayed by $t_2$ (waiting time), as shown in Fig. \ref{fig5: 2D spectroscpy}a. In addition to achieving spectral resolution along the probe energy (using a spectrometer like in standard pump-probe setups), 2DES adds spectral resolution along the excitation energy. This is done by measuring, at a fixed $t_2$, the transient reflectivity signal as a function of $t_1$ and then Fourier transforming it over $t_1$, resulting in a two-dimensional spectrum with probe and pump photon energies on the horizontal and vertical axes, respectively \cite{fuller2015experimental, tollerud2017coherent,collini20212d}.
The shape of the spectral features in the 2D map reflects the interactions with the environment and broadening mechanisms: peaks elongated along the diagonal direction indicate inhomogeneous broadening, with the anti-diagonal width representing the homogeneous broadening contribution, which is directly related to the decoherence time \cite{hamm2011concepts}. The ability to disentangle homogeneous from inhomogeneous broadening mechanisms is unique of 2DES and is unattainable in linear or pump-probe techniques because of the lack of spectral resolution along the excitation energy axis.  In the partially collinear configuration of 2DES (collinear pump pulses and a non-collinear probe), conventional pump-probe spectroscopy coincides to the case where $t_1$ is fixed to $t_1 = 0$. The measured pump-probe spectrum then corresponds to the integral of the 2D map along the excitation energy axis, if broadband pump pulses are employed, or to a horizontal slice of 2D spectrum in the case of narrowband excitation, thus preventing the resolution of the anti-diagonal width of the peaks.

Here, we performed 2DES on \ch{LaVO_3} thin film by employing degenerate and cross-polarized pump and probe pulses, of $\sim 30$~fs time duration, within the 1.45-1.9~eV spectral range, thus allowing to investigate the response of the Hubbard exciton. The experiment is performed in partially collinear scheme and reflection geometry \cite{milloch_fapi}, as detailed in Appendix D and sketched in Fig. S5 of the Supplemental Material. The sample is cooled down to $ T \lesssim T_c$, where the pump-probe transient reflectivity displays the maximum signal in the slow build-up component, and the 2D spectrum is then collected at selected $t_2$ delays. To partially mitigate the influence of the spectral shape of the light source on the 2D map \cite{camargo2017resolving}, we normalize the 2DES signal over the light spectrum along both pump and probe energy axes. Fig. \ref{fig5: 2D spectroscpy}b reports the 2D data measured for $T$ = 140 K and $t_2$~=~40~ps, which corresponds to the maximum of the pump-probe signal, and reveals a broad transient reflectivity response located around the diagonal of the 2D spectrum around 1.8~eV. 

In order to study the dynamics of the homogeneous linewidth of the excitonic resonance, we extract a line-cut of the 2D spectrum along the anti-diagonal direction, as displayed by the black dashed line in Fig. \ref{fig5: 2D spectroscpy}b, for different $t_2$ delays over a 100 ps window. In Fig. \ref{fig5: 2D spectroscpy}c the resulting signal is plotted along the anti-diagonal energy axis for a short time delay ($t_2$~=~100~fs) and at a later delay ($t_2$~=~40~ps). Normalization of the two anti-diagonal profiles for comparison purposes clearly shows a broader peak at $t_2$~=~40~ps as compared to the linewidth at ultrashort times. The full width at half maximum $\Gamma_{hom}$ is extracted for the five $t_2$ delays where 2DES is performed, which are highlighted by the colored dots plotted along the pump-probe dynamics in Fig. \ref{fig5: 2D spectroscpy}d, top panel. Since the width of the excitonic resonance is comparable to the pulse bandwidth, the extraction of absolute values for the homogeneous linewidth is difficult and subject to possible artifacts originating from the spectral shape of the pulse \cite{camargo2017resolving}. We therefore focus only on relative variations of the spectral response as a function of time delay $t_2$. The bottom panel in Fig. \ref{fig5: 2D spectroscpy}d reports the difference,  as a function of time delay $t_2$, between the linewidth $\Gamma_{hom,0}(t_2)$ and $\Gamma_{hom,0}(t_2=100~\text{fs})$, normalized to $\Gamma_{hom,0}(t_2=100~\text{fs})$, revealing a 20$\%$ broadening building up within tens of picoseconds. 

Similarly to pump-probe spectroscopy, 2DES shows a broadening of the exciton linewidth, whose dynamics is also compatible with the time evolution of $\Gamma_{HE}$ found in Sec. \ref{sec:pump-probe lvo}. 2DES gives additional insight, revealing that this broadening affects the homogeneous component of the linewidth and, therefore, indicates a decrease of the Hubbard exciton decoherence time due to disruption of the ordered background caused by the pump excitation. 
The $\Delta\Gamma_{hom}/\Gamma_{hom,0}$ obtained from 2DES ($\sim 20\%$) is larger than what has been measured in pump-probe ($\sim 0.5\%$, see Sec. \ref{sec:pump-probe lvo}) because, in the former case the linewidth variation is observed specifically along the anti-diagonal direction (homogeneous component), whereas the latter approach measures the total linewidth change from the signal projected onto the probe energy axis.
A direct comparison based on simulated 2D spectra shows that these different observables naturally yield markedly different relative broadenings; when the different excitation conditions and temperatures of the two experiments are further taken into account, the results are found to be quantitatively consistent (see Supplemental Material Sec. S5 for a detailed analysis).
The significant difference in the relative linewidth variation values obtained from the two experimental techniques suggests a strong inhomogeneous broadening,  which likely cannot be fully appreciated in the 2DES data presented here due to the use of light pulses with a spectrum comparable to the HE linewidth.

The observation that the time-resolved signal measured in ultrafast spectroscopy originates from a perturbation of the exciton decoherence time (homogeneous linewidth) justifies assumption (1), made in Sec. \ref{sec: gl}, about the correspondence of $\Gamma_{HE}$ to the exciton lifetime. In the next section, we address assumption (2) and discuss the link between the exciton linewidth and the order parameter.

\section{Discussion}
\label{sec: dmft}
In order to connect the experimental results to a microscopic description of the electronic properties of the material and investigate the effect of the establishment of long-range orders below the critical temperature,  we performed DMFT calculations starting from the simplest model that captures the symmetry-breaking transition in \ch{LaVO3}. DMFT is an accurate many-body method which includes non-perturbatively correlation effects and allows for calculations of experimentally accessible spectra \cite{Georges1996}. Here, the DMFT calculations have been performed for a three-orbital model, based on density-functional theory bandstructure. The bandstructure has been computed using Quantum Espresso \cite{QE} and maximally localized Wannier orbitals have been derived using Wannier90 \cite{Wannier90}. Only the three low-lying orbitals of vanadium are included, as they are those that mainly contribute to the relevant bands. The interactions are then included in the popular density-density Hubbard-Kanamori form 
\begin{align}
\label{Hamiltonian}
H_{int} & = U\sum_{i,a} n_{ia\uparrow}n_{ia\downarrow} 
+ (U-3J)\sum_{i,a<b, \, \sigma} n_{ia\sigma}n_{ib\sigma} \nonumber\\
& + (U-2J)\sum_{i,a\neq b} n_{ia\uparrow}n_{ib\downarrow}  
\end{align}
where $i$, $a, b =1,2, 3$, $\sigma$ are respectively site, orbital and spin indices. 
The (screened) interaction parameters appearing in Eq. \ref{Hamiltonian} are set to  $U=5$~eV and $J =0.68$~eV, following previous literature \cite{PhysRevLett.92.176403,PhysRevB.54.5368,PhysRevB.106.115110}.
We performed DMFT calculations using a finite-temperature exact diagonalization (ED) solver \cite{Capone} using a scalar Lanczos algorithm and bath representations analogous to Ref. \citenum{EDIpack}, with three levels in the bath allowing for spontaneous symmetry breaking both in the orbital and magnetic sections. The number of states kept in the finite-temperature averages has been taken as 40.
In agreement with Ref. \citenum{PhysRevB.106.115110}, the low-temperature solution has both G-type orbital ordering and C-type AFM ordering. Within the accuracy of our calculation, magnetic ordering appears to take place at a  slightly larger temperature than orbital ordering. However, we consider this difference below the resolution of our approach, in light of the reduced accuracy of our method when the temperature is increased. ìTherefore, in the following, we will contrast a low-temperature ordered phase with a high-temperature disordered phase, which has neither magnetic nor orbital long-range order. The theoretical transition temperatures ($T_{c}^{DMFT} \simeq$ 290 K) are nonetheless significantly larger than the experimental ones due to the mean-field nature of DMFT.

\begin{figure}[]
\centering
\includegraphics[width=8.5cm]{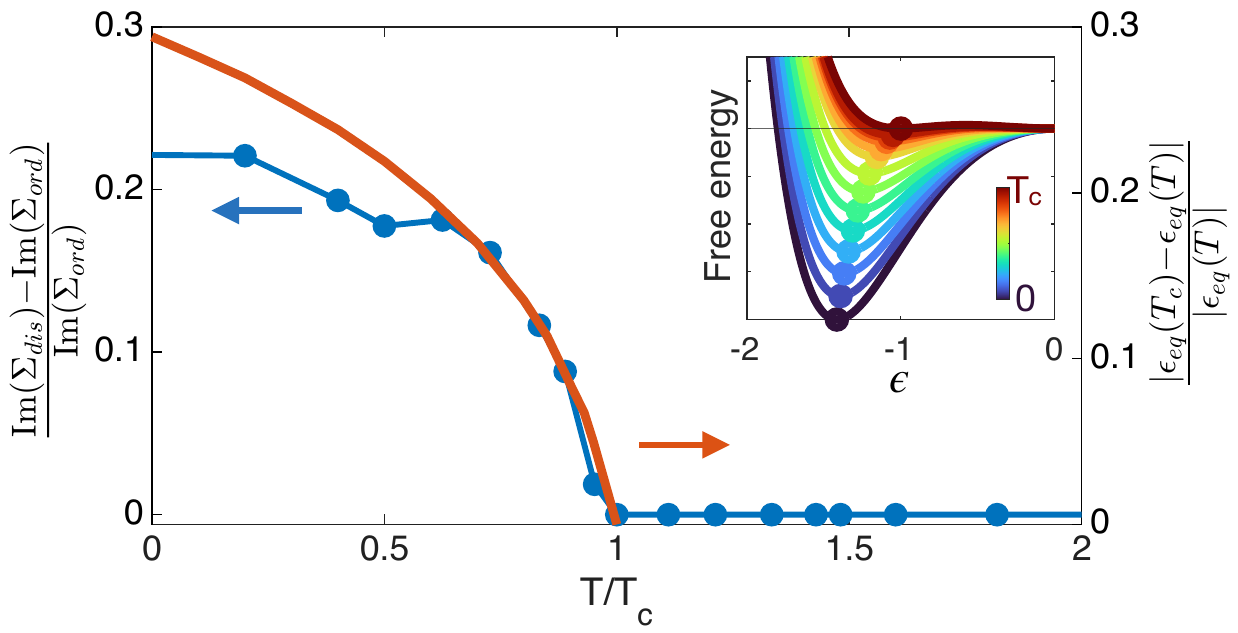}
\caption{Blue markers (left $y$-axis) show the temperature dependence of the variation in the scattering rate - obtained by DFMT calculation as $\text{Im}(\Sigma)$ (Fig. S9 Supplemental Material) - upon suppression of spin and orbital orders.  The red solid line (right $y$-axis) shows the temperature-dependence of the order parameter $\epsilon_{eq}$, estimated as the position of the minimum of the free energy in Eq. \ref{eq: free energy eq} with $\alpha = 5$, plotted in the inset.}
\label{fig: dmft order param}
\end{figure}

DMFT results are then used to compute the scattering rate of the Hubbard exciton as the imaginary part of the self-energy, $\Sigma$ \cite{giannetti2016}, with the aim of comparing the theoretical model with the time-resolved experiments and the ultrafast suppression of spin and orbital orders. 
$\text{Im}(\Sigma)$ is, therefore, extracted for two different solutions: the most stable one, representing the ordered case where spin and orbital orders are established below $T_c$, and a disordered (paramagnetic) solution, where the long-range orders are suppressed. 
Specifically, from the DMFT calculation we extract an average of $\text{Im}(\Sigma)$ in a 0.2~eV window around the Hubbard exciton level on the real-frequency axis. This averaging is necessary to smooth out the discrete features inherent to the exact diagonalization solution. We note here that our DMFT calculations cannot reproduce the full nature of the Hubbard exciton, which requires the inclusion of non-local vertex corrections \cite{kim2018signatures}. Yet, our calculations show an essentially coherent peak close to the edge of the Hubbard band that we can safely associate with the Hubbard exciton.
 The values of $\text{Im}(\Sigma)$ obtained for the two cases are reported in Fig.~S9 Supplemental Material as a function of temperature. In Fig. \ref{fig: dmft order param} we instead display the difference between the $\text{Im}(\Sigma)$ values in the disordered (paramagnetic) case and that in the ordered case, normalized to the value of the ordered solution ($[{\mathrm{Im}(\Sigma_{dis})-\mathrm{Im}(\Sigma_{ord})}]/{\mathrm{Im}(\Sigma_{ord})}$). 
At fixed temperature, the disordered solution displays a significant increase in $\text{Im}(\Sigma)$, compared to the ordered case, for $T<T_c$ (see also Fig. S9 Supplemental Material). This behavior indicates that the Hubbard exciton scattering rate strongly depends on the presence of the long-range order. As discussed above, our calculations do not allow to easily disentangle the contribution of the magnetic and orbital ordering. Nonetheless, we report that, considering only the magnetic ordering, we obtain scattering rates which are reduced by 15-20\%, and show a temperature dependence similar to what is reported in Figs. \ref{fig: dmft order param} and S9, thus suggesting a major role played by the magnetic order.


The DMFT results presented here show that the exciton scattering rate is strongly affected by the establishment of long-range order, and can therefore be taken as representative of the order parameter, justifying assumption (2) made in Sec. \ref{sec: gl}. In order to show that the scattering rate can indeed map the order parameter, in Fig. \ref{fig: dmft order param} we compare the temperature-dependence of the scattering rate, obtained from DMFT, to that of the order parameter of the free energy $f$ that has been determined from the analysis of the pump-pump experiment (i.e. $f$ is given by Eq. \ref{eq: free energy eq} with $\alpha = 5$, as discussed in Sec. \ref{sec: gl}). Specifically, the red line in Fig. \ref{fig: dmft order param} reports $|(\epsilon_{eq}(T_c)-\epsilon_{eq}(T))/\epsilon_{eq}(T)|$, where $\epsilon_{eq}$ indicates the minimum of $f$, as plotted in the inset. We observe that the variation of DMFT scattering rate follows the trend of the Ginzburg-Landau order parameter, characterized by a decreasing behavior as the temperature is increased, until reaching zero at $T > T_c$. 

DMFT calculations overall confirm that the suppression of long-range order results in a decrease in intrinsic lifetime (increase in scattering rate), as observed experimentally for the Hubbard exciton upon pump-perturbation of the spin and orbital ordered background. This validates the ascription of the pump-probe and 2DES signals to an increase in the scattering rate that is due to the coupling with the spin and orbital long-range orders. We lastly note that the signal enhancement in proximity of $T_c$ observed in ultrafast spectroscopy is not directly related to the temperature dependence of scattering rate variation due to suppression of the long-range order ($[{\mathrm{Im}(\Sigma_{dis})-\mathrm{Im}(\Sigma_{ord})}]$ in Fig. \ref{fig: dmft order param}); rather, it is the coupling to the light-induced electronic excitations that becomes more efficient in proximity of $T_c$ and leads to a larger variation of the order parameter upon perturbation, in agreement with the curve in Fig. \ref{fig: dmft order param} being steeper right below $T_c$.

\section{Conclusions}
In this work, we analyzed pump-probe and 2DES data in a \ch{LaVO3} thin film. The use of a broadband probe in pump-probe spectroscopy revealed a broadening in the Hubbard exciton linewidth upon pump excitation. 2DES further showed that, due to the pump excitation, there actually is an increase in the homogeneous component of the linewidth, $\Gamma_{hom}$, which can be related to the order parameter of the spin and orbital degrees of freedom of the system, as shown by DMFT calculation for a three-band model of \ch{LaVO3}.
When light pulses are used to perturb the system, the photoinduced electronic excitations couple to the order parameter and result in a disruption of the spin/orbital ordered background. This coupling and the resulting dynamics of the order parameter are captured within Ginzburg-Landau theory, which explains the signal enhancement observed in proximity of $T_c$ and the critical slowing down of the order parameter dynamics upon approaching the transition temperature. 


%

Overall, our results represent a direct observation of the decoherence time of the Hubbard exciton being strongly affected by the coupling with long-range orders. Although this was here observed specifically on \ch{LaVO3}, the change in scattering rate upon crossing a symmetry-breaking phase transition is likely a general mechanism that extends to many other strongly correlated materials displaying ordered phases (e.g. magnetic or charge order). Some examples include, for instance, charge order and antiferromagnetic correlations in copper oxides (e.g. \ch{Nd_{2-x}Ce_xCuO_4}) \cite{da2015charge,frano2020charge}, and charge density wave (e.g. 1T-\ch{TaS2}) \cite{Lee2019}. The variation of the fluctuations of the thermal bath, along with their interaction with electronic excitations, in proximity of a phase transition, could provide a novel mechanism to tune the ultrafast electronic decoherence dynamics in solids. 

\section*{Acknowledgments}
A.M., P.F. and C.G. acknowledge financial support from MIUR through the PRIN 2017 (Prot. 20172H2SC4 005) and PRIN 2020 (Prot. 2020JLZ52N 003) pprograms and from the European Union - Next Generation EU through the MUR-PRIN2022 (Prot. 20228YCYY7) program. Y.L. and K.M.S. would like to acknowledge support from NSF DMR-2104427. This work made use of the thin film facility of the Platform for the Accelerated Realization, Analysis, and Discovery of Interface Materials (PARADIM), which is supported by the NSF under Cooperative Agreement No. DMR-2039380.

\section*{Appendix: Materials and methods}

\subsection{Samples}  
The ultrafast spectroscopy data analyzed here are collected from a 30~nm thick \ch{LaVO_3} thin film sample grown on LSAT (\ch{(LaAlO_3)_{0.3}(Sr_2TaAlO_6)_{0.7}}) substrate by Molecular Beam Epitaxy (MBE). Similar results have also been obtained for 150~nm \ch{LaVO_3} thin film grown by Pulsed Laser Deposition (PLD) on \ch{SrTiO3} substrate.
For ultrafast spectroscopy measurements, the samples were mounted inside a closed-cycle helium cryostat. \\

\subsection{Pump-probe spectroscopy setup}
The time-resolved transient reflectivity setup is based on a Yb:KGW laser system (Pharos, Light Conversion) emitting 300~fs pulses at 1030~nm. A portion of the laser output pumps an optical parametric amplifier (Orpheus-F, Light Conversion) whose signal at 880~nm (time duration $\sim 50$ fs) is used as the excitation pulse in the experiment; the rest of the laser light is employed to synthesize a supercontinuum probe pulse, obtained by means of White Light Generation in a 6-mm-thick YAG crystal. A linearly motorized stage delays the pump pulse in a time window covering $\sim$~330~ps. The pump beam is focused to a 160 \textmu m $\times$ 180 \textmu m spot size, being $\approx$ 2.5 times larger than the probe spot size at the sample position (65 \textmu m $\times$ 65 \textmu m). Spectral resolution over the probe energy axis is achieved by employment of a common-path birefringent interferometer (GEMINI by NIREOS) and Fourier transform of the generated interferogram \cite{preda2016broadband}. A mechanical chopper working at 2.6~kHz modulates the pump beam in order to perform lock-in acquisition of the transient reflectivity signal. The laser repetition rate employed in the pump-probe measurements presented here is 400~kHz; no change in the sample response is observed upon decreasing the repetition rate while keeping fixed the energy per pulse of pump and probe beams. \\

\subsection{Pump-probe data analysis}
In order to relate the pump-probe signal to changes in the optical properties of the sample, we analyze the broadband transient reflectivity data by fitting the vertical slices of the pump-probe map to a differential reflectivity $(R^{neq}-R^{eq})/R^{eq} $. The differential fit is performed, for all time delays $t_2 \geq $ 500~fs, as described below. 

We start from a parametrization of the equilibrium optical properties of \ch{LaVO3} - which are reported in Ref. \citenum{miyasaka2002} and plotted in Fig. \ref{fig1: LaVO3 intro}b - based on a multi-peak Drude-Lorentz model. The dielectric function $\varepsilon$ as a function of frequency $\omega$ is given by
\begin{equation}
     \varepsilon(\omega) = \varepsilon_{\infty} +\sum_j \frac{\omega_{p,j}^2}{\omega_{0,j}^2-\omega^2-i\Gamma_j\omega}.
     \label{eq: DL model}
 \end{equation}
where $\varepsilon_{\infty}$ is the value of the dielectric constant at high frequency, $\omega_{p,j}$ is the plasma frequency, $\omega_{0,j}$ is the central frequency of the optical transition and $\Gamma_j$ is the linewidth of the $j$-th oscillator. The real part of the optical conductivity $\sigma_1$ is then related to $\varepsilon$ according to 
\begin{equation}
\sigma_1 = \frac{\omega \varepsilon_2}{4 \pi}
\label{eq: DL model sigma}
\end{equation}
with $\varepsilon_2$ indicating the imaginary part of Eq. \ref{eq: DL model}. Eq. \ref{eq: DL model sigma} with three Drude-Lorentz oscillators ($j = 1,2,3$) is used to fit the optical conductivity data in Fig. \ref{fig1: LaVO3 intro}b. The lowest energy oscillator, which is highlighted by the filled area in Fig. \ref{fig1: LaVO3 intro}b, is related to the Hubbard exciton. It is centered at $\omega_{0,HE}=1.82$~eV and has a spectral weight ($\propto \omega_{p,HE}^2$) that increases as the temperature is lowered below $T_c$, from $\omega_{p,HE} = 0.9$~eV at room temperature to 1.6~eV at cryogenic temperature. Since this parametrization is based on literature data measured on bulk \ch{LaVO3} rather than thin film, and the differential fit does not strongly depend on the specific value used for the equilibrium HE plasma frequency (as we are looking at the variation in the optical properties), we consider the same equilibrium parameters to fit both measurements at 40~K and 110~K. The parameters adopted to reproduce the equilibrium dielectric function of low-temperature \ch{LaVO3}, employed in the differential fitting of the pump-probe data, are reported in Table \ref{tab:lvo eq param}. 

\begingroup
\renewcommand*{\arraystretch}{1.5}
\begin{table}[h]
\begin{center}
\begin{tabular}{ |c|c|c|c| } 
 \hline
 & $\omega_{0,j}$ (eV) & $\omega_{p,j}$ (eV) & $\Gamma_{j}$ (eV) \\  \hhline{|=|=|=|=|}
 $j = 1$ (HE) & 1.82 & 1.31 & 0.56 \\ 
 $j = 2$ & 2.40 & 1.68 & 1.21 \\ 
 $j = 3$ & 4.61 & 2.16 & 0.70 \\ 
 \hline
\end{tabular}
\end{center}
\caption{Parameters of the \ch{LaVO3} low-temperature equilibrium dielectric function (Eq. \ref{eq: DL model} with $\varepsilon_{\infty} = 3$) employed in the differential fit of the pump-probe data. }
\label{tab:lvo eq param}
\end{table}
\endgroup

The pump-probe signal can then be described by pump-induced changes in optical properties of the sample, captured through the variation of some of the parameters modeling $\varepsilon$ in Eq. \ref{eq: DL model}, as compared to equilibrium. We specifically find that, in order to fit the transient reflectivity data of Fig. \ref{fig3: dRR_maps}, two parameters related to the HE need to be varied in the out-of-equilibrium configuration, namely the plasma frequency $\omega_{p,HE}$ and the width $\Gamma_{HE}$. The transient reflectivity data $\Delta R /R$ at fixed $\Delta t$ are therefore fitted (see red and blue lines in Fig. \ref{fig3: dRR_maps}b and d) according to the following equation: 
 \begin{equation}
     \frac{R(\omega_{p,HE}^{neq},\Gamma_{HE}^{neq})-R(\omega_{p,HE}^{eq},\Gamma_{HE}^{eq})}{R(\omega_{p,HE}^{eq},\Gamma_{HE}^{eq})}
     \label{eq: spectra fit}
 \end{equation}
where the reflectivity $R$ is obtained by means of transfer matrix formalism for a \ch{LaVO3} 30~nm film (whose dielectric function is given by Eq. \ref{eq: DL model} with parameters in Table \ref{tab:lvo eq param}) on LSAT substrate. $neq$ and $eq$ superscripts indicate the plasma frequency or exciton width parameters in out-of-equilibrium and equilibrium conditions, respectively.  

The fitting results reveal that, in order to account for the spectral response in $\Delta R/R$, two effects of the HE resonance need to be included: i) a decrease of excitonic spectral weight, ii) a broadening of the HE peak linewidth $\Gamma_{HE}$, as compared to equilibrium. \\

\subsection{2DES setup}
A sketch of the scheme employed for 2DES is reported in Fig. S5 Supplemental Material. Both pump and probe pulses are generated by a home-built non-collinear optical parametric amplifier (NOPA), seeded by a Yb:KGW laser (Pharos by Light Conversion). The NOPA signal covers the spectral range between 1.45 and 1.9~eV and is compressed to $\sim$30~fs time duration (see Fig. S5 in Supplemental Material) by multiple bounces on a pair of chirped mirrors. A beam splitter separates the light into pump and probe; the pump beam then goes through a common-path birefringent interferometer (GEMINI 2D by NIREOS) that generates the pair of phase-coherent excitation pulses \cite{brida2012phase}. Two separate chirped mirrors compressors are employed on the pump and probe beams to compensate for the additional dispersion introduced by various optical elements: GEMINI 2D, beam splitter, half-waveplate and cryostat window for the pump, a half-waveplate and the cryostat window for the probe. In order to minimize pump scattering, pump and probe pulses are orthogonally polarized. They are then focused onto the sample by two concave mirrors; the focused spot size is 200 \textmu m $\times$ 110 \textmu m for the pump beam and 70 \textmu m  $\times$ 60 \textmu m for the probe beam. The pump-probe time delay $t_2$ is controlled through a linearly motorized stage and is scanned over a 100 ps time window.  The delay between the two pump pulses, $t_1$, is controlled by varying the beam insertion of one of the GEMINI 2D birefringent wedges and it is continuously scanned between -35~fs and 70~fs for each measured $t_2$. The 2D signal propagates collinearly with the probe beam and is collected in reflection geometry. At each ($t_1$,$t_2$) the spectrally resolved signal is measured with the same setup described for pump-probe experiment, which employs GEMINI interferometer and lock-in acquisition. For the 2DES measurements presented in this work, the laser repetition rate was set to 40~kHz.

\bibliography{Refs}
\end{document}


\title{Supplemental Material for \\ Tracking the local order parameter through the Hubbard exciton decoherence time in the Mott-Hubbard insulator \ch{LaVO3}}

\author{Alessandra Milloch}
\email[]{alessandra.milloch@unicatt.it}
\affiliation{Department of Mathematics and Physics, Università Cattolica del Sacro Cuore, Brescia I-25133, Italy}
\affiliation{ILAMP (Interdisciplinary Laboratories for Advanced
Materials Physics), Università Cattolica del Sacro Cuore, Brescia I-25133, Italy}
\affiliation{Department of Physics and Astronomy, KU Leuven, B-3001 Leuven, Belgium}

\author{Paolo Franceschini}
\affiliation{CNR-INO (National Institute of Optics), via Branze 45, 25123 Brescia, Italy}
\affiliation{Department of Information Engineering, University of Brescia, Brescia I-25123, Italy}

\author{Pablo Villar-Arribi}
\affiliation{International School for Advanced Studies (SISSA), Via Bonomea 265, Trieste 34136, Italy}

\author{Sandeep Kumar Chaluvadi}
\affiliation{CNR-IOM (Istituto Officina dei Materiali) S.S. 14, km 163.5, I-34149 Trieste, Italy}

\author{Pasquale Orgiani}
\affiliation{CNR-IOM (Istituto Officina dei Materiali) S.S. 14, km 163.5, I-34149 Trieste, Italy}

\author{Giancarlo Panaccione}
\affiliation{CNR-IOM (Istituto Officina dei Materiali) S.S. 14, km 163.5, I-34149 Trieste, Italy}

\author{Giorgio Rossi}
\affiliation{CNR-IOM (Istituto Officina dei Materiali) S.S. 14, km 163.5, I-34149 Trieste, Italy}
\affiliation{Dipartimento di Fisica, Università degli Studi di Milano, Via Celoria 16, 20133 Milano, Italy}

\author{Yang Liu}
\affiliation{Laboratory of Atomic and Solid State Physics, Cornell University, Ithaca, NY 14853, USA}

\author{Darrell G. Schlom}
\affiliation{Department of Materials Science and Engineering, Cornell University, Ithaca, New York 14853, USA}
\affiliation{Kavli Institute at Cornell for Nanoscale Science, Ithaca, New York 14853, USA}
\affiliation{Leibniz-Institut für Kristallzüchtung, Max-Born-Str. 2, 12489 Berlin, Germany}

\author{Kyle M. Shen}
\affiliation{Laboratory of Atomic and Solid State Physics, Cornell University, Ithaca, NY 14853, USA}
\affiliation{Kavli Institute at Cornell for Nanoscale Science, Ithaca, New York 14853, USA}

\author{Massimo Capone}
\affiliation{International School for Advanced Studies (SISSA), Via Bonomea 265, Trieste 34136, Italy}
\affiliation{CNR-IOM (Istituto Officina dei Materiali), Via Bonomea 265, Trieste 34136, Italy}

\author{Claudio Giannetti}
\email[]{claudio.giannetti@unicatt.it}
\affiliation{Department of Mathematics and Physics, Università Cattolica del Sacro Cuore, Brescia I-25133, Italy}
\affiliation{ILAMP (Interdisciplinary Laboratories for Advanced
Materials Physics), Università Cattolica del Sacro Cuore, Brescia I-25133, Italy}
\affiliation{CNR-INO (National Institute of Optics), via Branze 45, 25123 Brescia, Italy}

\maketitle

\section{Fluence-dependent pump-proby dynamics}
\begin{figure*}[h!]
\centering
\includegraphics[width=13cm]{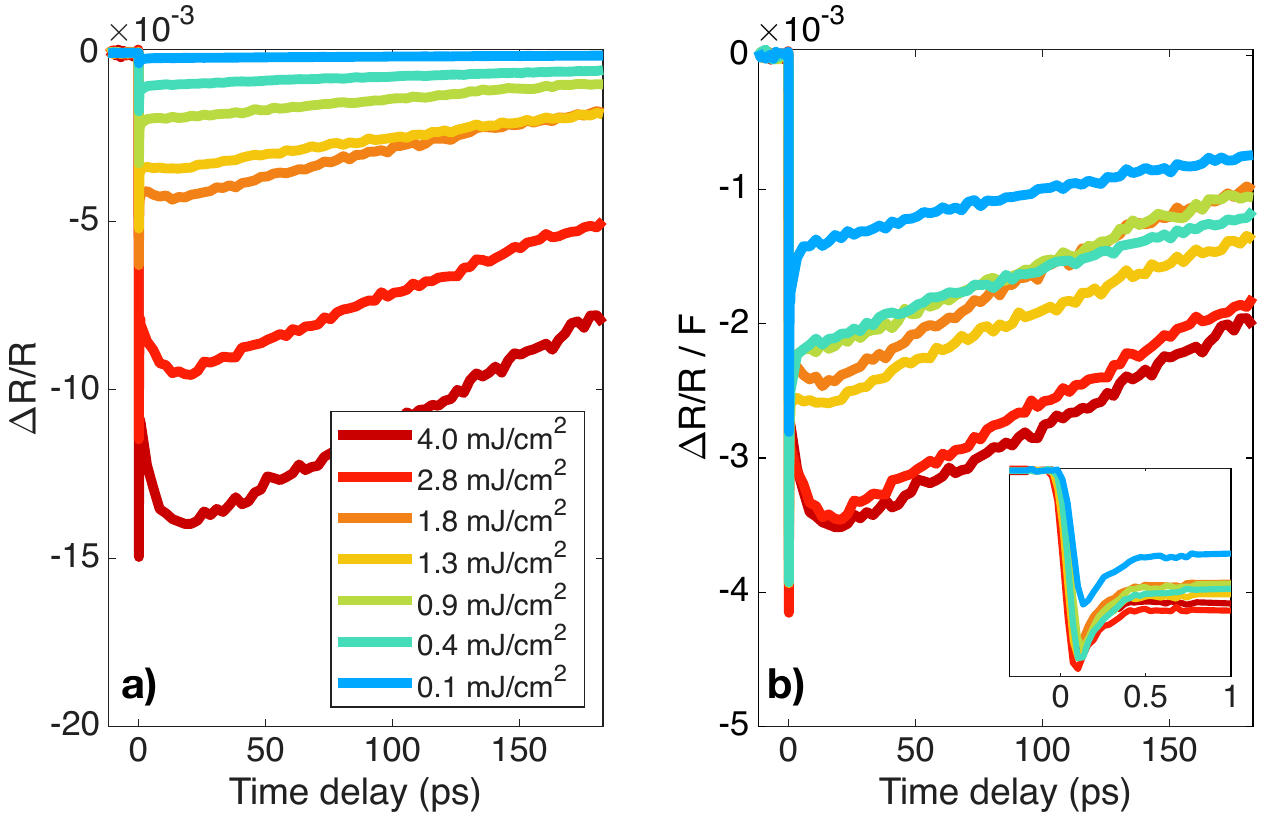}
\caption{a) Fluence dependence of the transient reflectivity signal, measured at 70 K, 1.65 eV pump, 1.77 eV probe photon energy. b) Same pump-probe traces plotted in a) normalied by the fluence $F$ for comparison purposes. Upon normalization, all traces collapse onto each other at early times (0-0.5 ps, see inset), demonstrating that the ultrafast response on the femtosecond timescale scales linearly with fluence. The slight deviation of the lowest-fluence trace at short delays is likely due to uncertainties in the measurement of the incident power at low laser intensities. In contrast, the slower component (1-200 ps) shows a clear dependence on excitation intensity, with an amplitude that is enhanced and increases non-linearly at higher fluence. }
\label{fig: fluence}
\end{figure*}

\clearpage
\section{Ginzburg-Landau model: temperature and fluence dependence}
\begin{figure*}[h!]
\centering
\includegraphics[width=15cm]{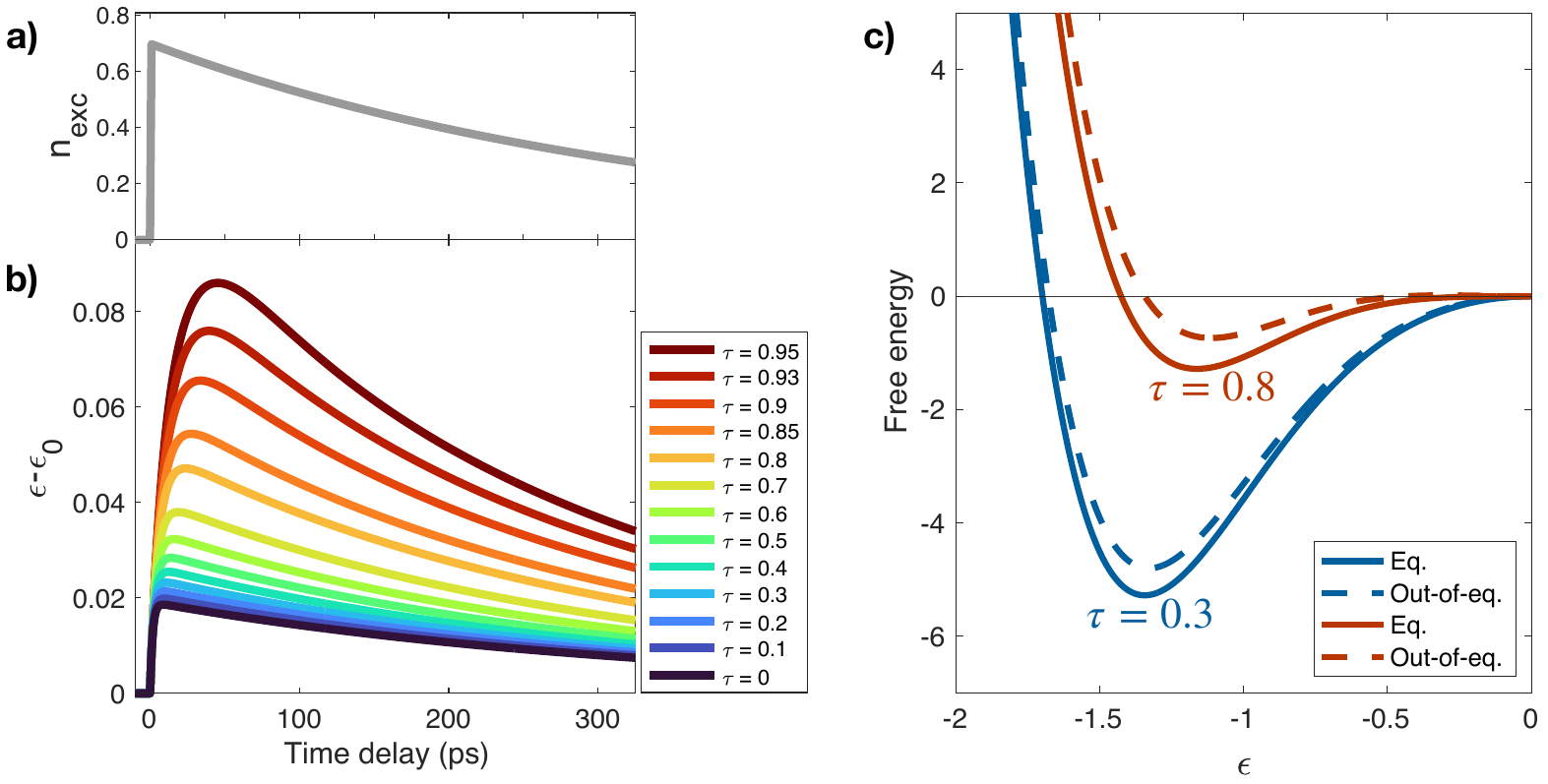}
\caption{a) Example of dynamics of $n_{exc}$ (Eq. 1) b) Order parameter dynamics obtained from the kinetic equation (Eq. 4) for several normalized temperatures $\tau$. The dynamics is initiated by $n_{exc}$ that couples to the orbital/spin order and perturbs the free energy. The parameters used in the time evolution simulation plotted here are: $ \alpha = 5$, $g = 0.6$, $n_{exc} = 0.7$, $\gamma = 0.01~ps^{-1}$. c)  Free energy functional at equilibrium (solid lines) and right after excitation of $n_{exc}$ (dashed lines) computed according to Eq. 2 and 3, respectively, using the same parameters of the simulations in panels a and b. }
\label{fig: GL simul}
\end{figure*}

\begin{figure*}[h!]
\centering
\includegraphics[width=10cm]{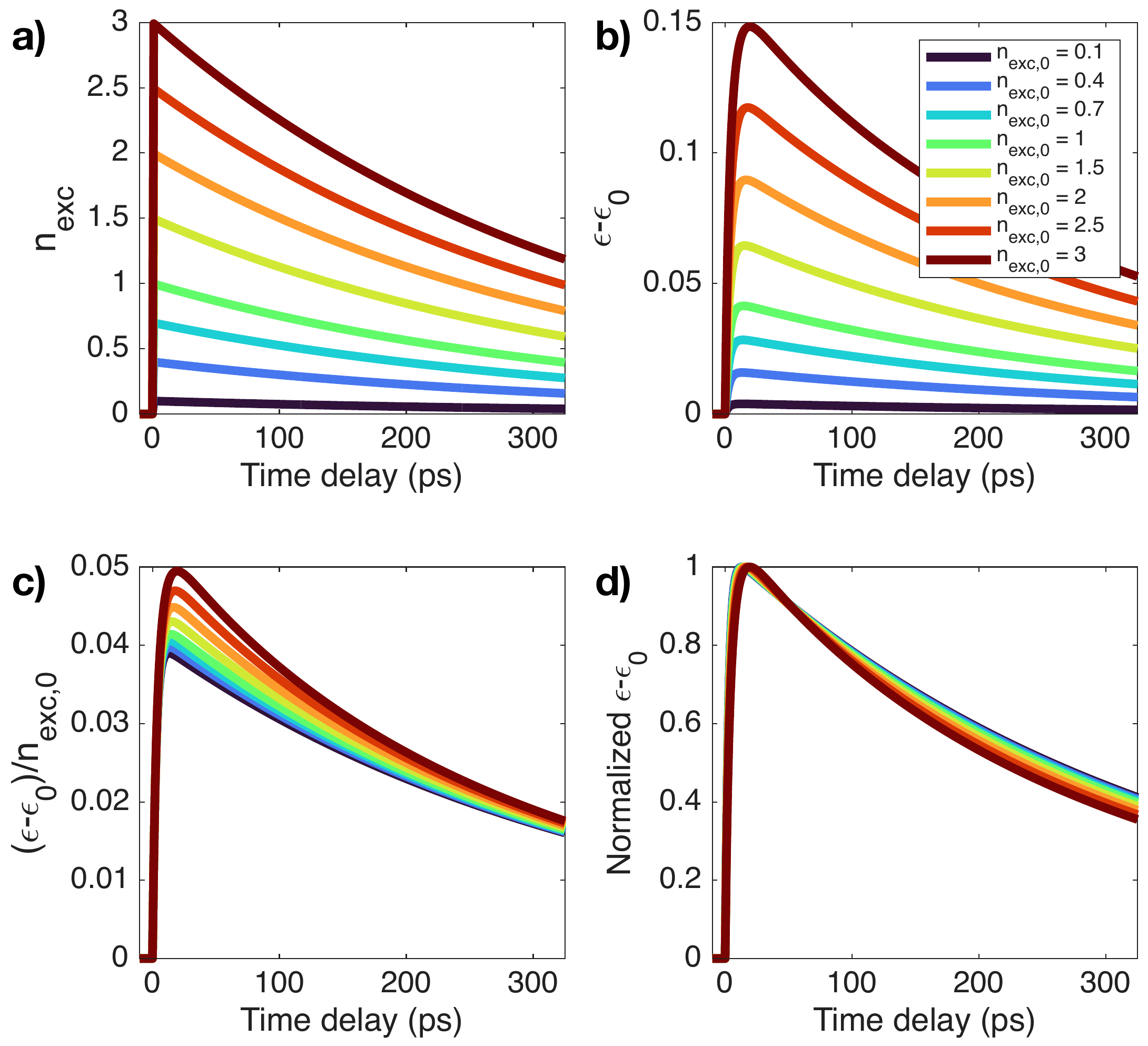}
\caption{Ginzburg-Landau model simulations of the fluence dependence of the order-parameter dynamics. a) Excitation strength parameterized by the amplitude of $n_{exc}$. b) Resulting order parameter dynamics, showing an increasing response with excitation intensity. c) Same order-parameter dynamics as in panel b), normalized by the excitation intensity, highlighting the nonlinear scaling of the slow-component amplitude. d) Same order-parameter dynamics as in panel b), normalized to unity, illustrating the weak dependence of the buildup and decay times on fluence.}
\label{fig: GL simul fluence}
\end{figure*}

\section{Ginzburg-Landau model with time-dependent temperature}
\begin{figure*}[h!]
\includegraphics[width=18cm]{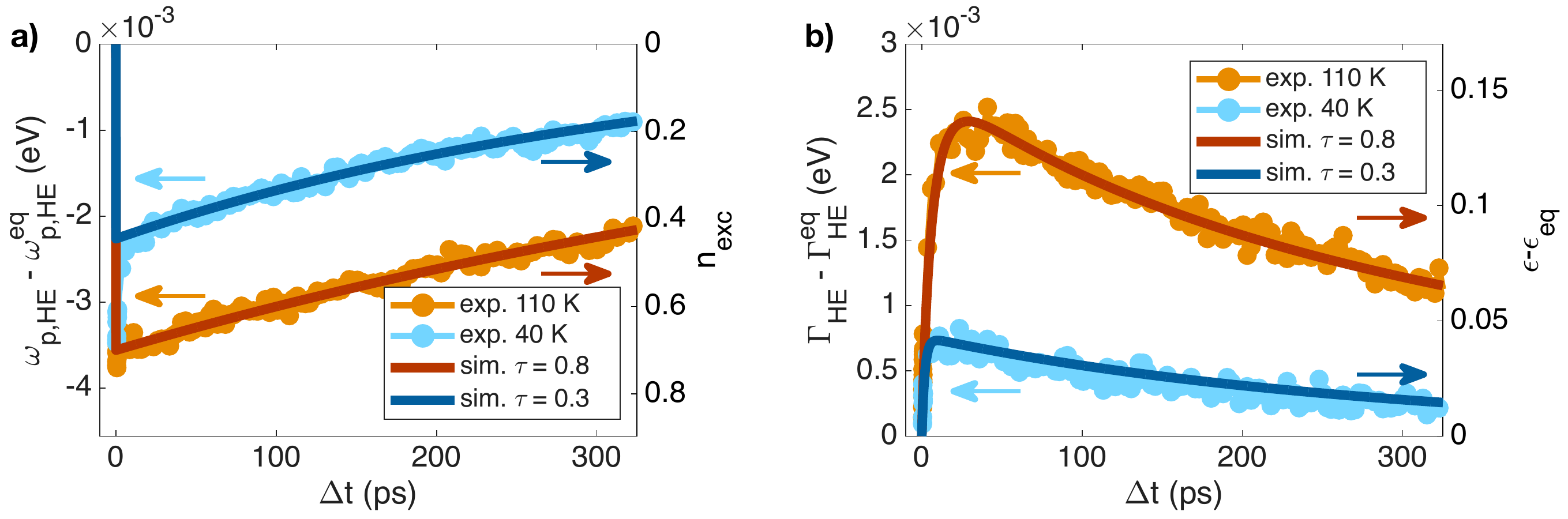}
\caption{Results of the refined Ginzburg-Landau model considering a time-dependent temperature (Eq. \ref{eq: time-dep temp}) with $T_\tau$ = 250 ps, $\Delta \tau$ = 0.05, $g$ = 1.1,  $\alpha $= 5, $\gamma$ = 0.012 ps$^{-1}$.}
\label{fig: refined GL}
\end{figure*}

The experimental relaxation of the HE linewidth on the hundreds-of-picoseconds timescale (Fig. 3b) is observed to be faster than that of the plasma frequency (Fig. 3a), suggesting the presence of an additional decay channel directly affecting the order-parameter dynamics, beyond that associated with $\tau_{exc}$. 
A natural origin of this behavior could be attributed to laser-induced heating: following absorption of the pump pulse, the system temperature increases on ultrafast timescales and subsequently relaxes back toward equilibrium. To account for this effect, we consider an extension of the Ginzburg-Landau framework with a time-dependent temperature, as a complementary refinement of the model presented in the main text. Specifically, we introduce a time-dependence for the normalized temperature $\tau(t)$ entering in Eq. 3, modeled as:
\begin{equation}
   \tau(t) = \tau_0 + \Delta \tau e^{-t/T_\tau},
   \label{eq: time-dep temp}
\end{equation}
where $T_{\tau}$ represents the temperature recovery time and $\tau_0$ is the initial temperature of the system. The results of the integration of the order parameter dynamics equation (Eq. 3 and 4) considering Eq. \ref{eq: time-dep temp}, are shown in Fig. \ref{fig: refined GL}. Using $T_{\tau}$=250 ps yields improved agreement between the simulated and experimental HE linewidth dynamics at long delays. \\
While this refined model provides a more accurate description of the recovery dynamics, the simpler Ginzburg-Landau formulation presented in the main text already captures the essential physics and relevant critical behavior, namely the emergence of the delayed response on the tens-of-picoseconds timescale and its enhancement with temperature and excitation fluence, while keeping the number of free parameters to a minimum and avoiding overfitting.

\clearpage
\section{2D spectroscopy setup}
\begin{figure*}[h!]
\includegraphics[width=16cm]{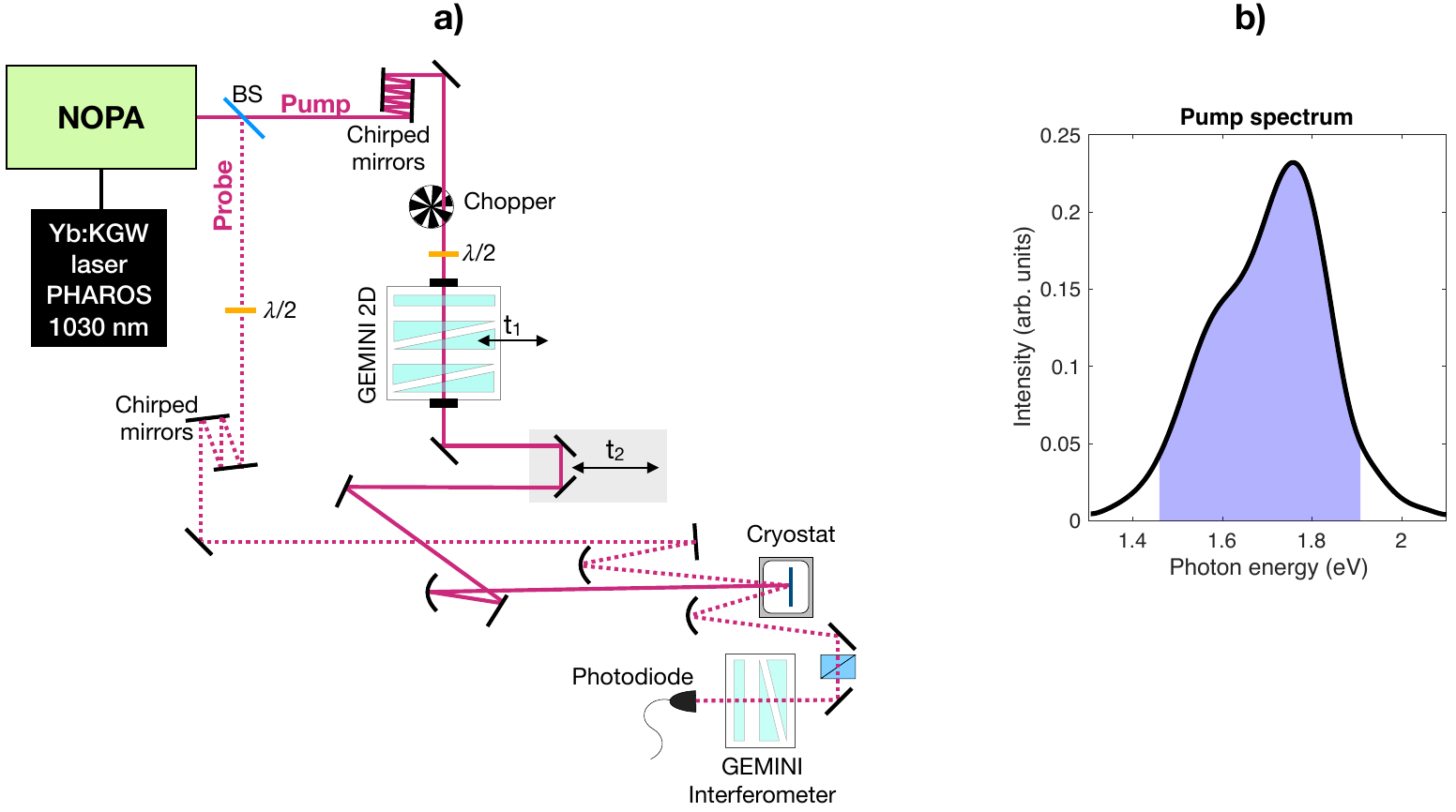}
\caption{a) Sketch of the 2DES experimental setup. b) Spectrum of the NOPA beam, employed for both pump and probe pulses. The shaded area represents the spectral window where 2D spectra are analyzed.}
\end{figure*}

\begin{figure*}[h!]
\includegraphics[width=15cm]{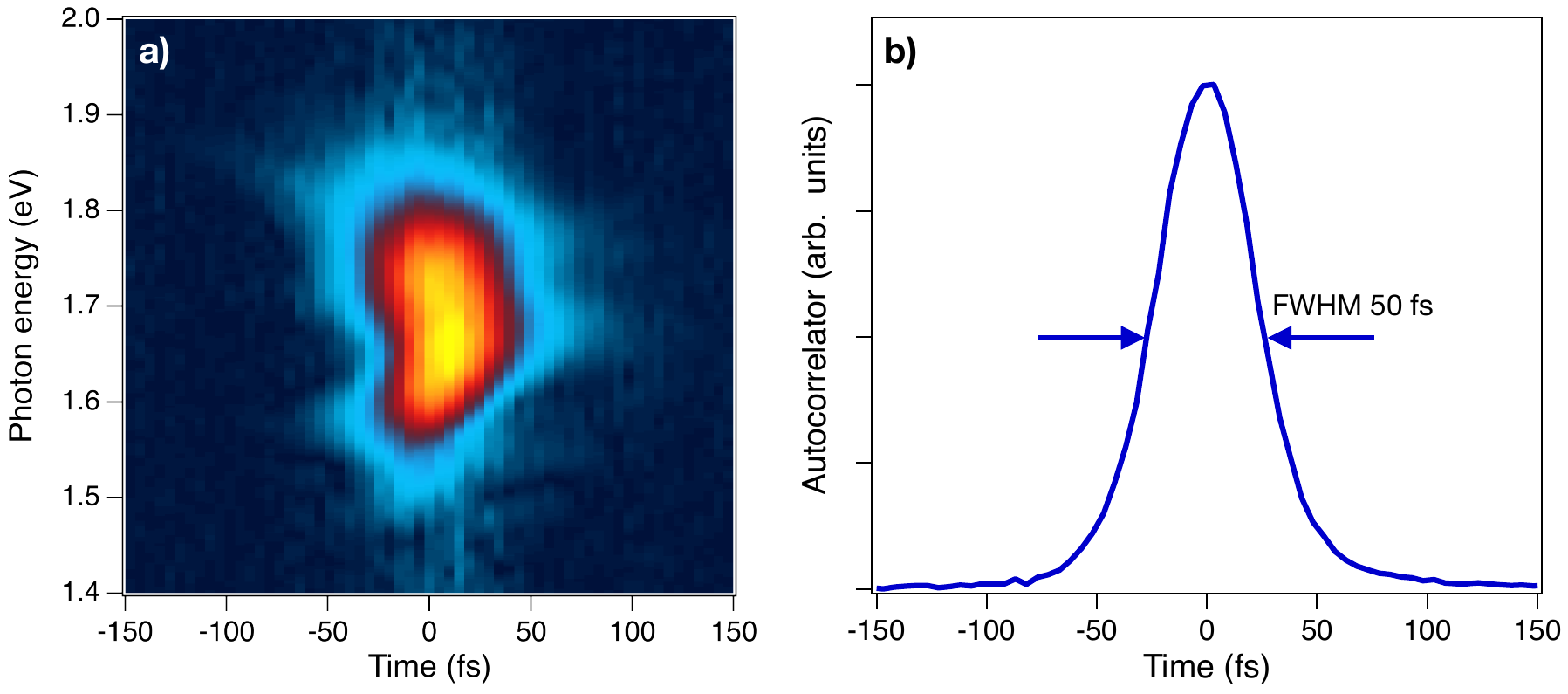}
\caption{a) PG-FROG (Polarization-Gating Frequency Resolved Optical Gating) trace of the NOPA signal used as pump and probe pulses of the 2DES experiment. b) Autocorrelator of the NOPA pulse, indicating a 35 fs time duration.}
\end{figure*}

\clearpage
\section{Linewidth broadening in 2DES and pump-probe experiments}

\begin{figure}[h!]
\centering
\includegraphics[width=16cm]{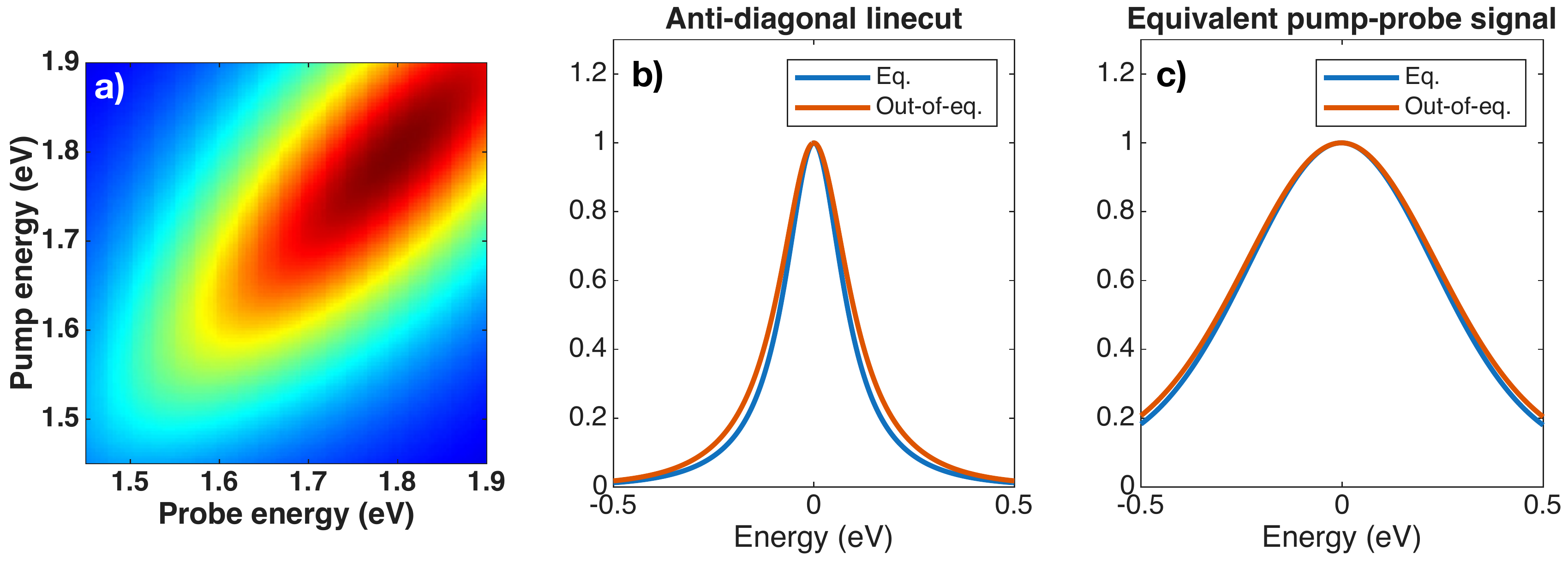}
\caption{a) Simulated 2DES spectrum for a two-level system with Gaussian inhomogeneous broadening. b) Anti-diagonal line cut of the 2DES spectrum, showing the linewidth change induced by a $20\%$ increase in the homogeneous dephasing rate $\Gamma$. c) Corresponding pump-probe spectrum obtained by projecting the 2DES signal onto the probe photon energy axis. }
\label{fig: broadening}
\end{figure}

\begin{figure}[h!]
\centering
\includegraphics[width=18cm]{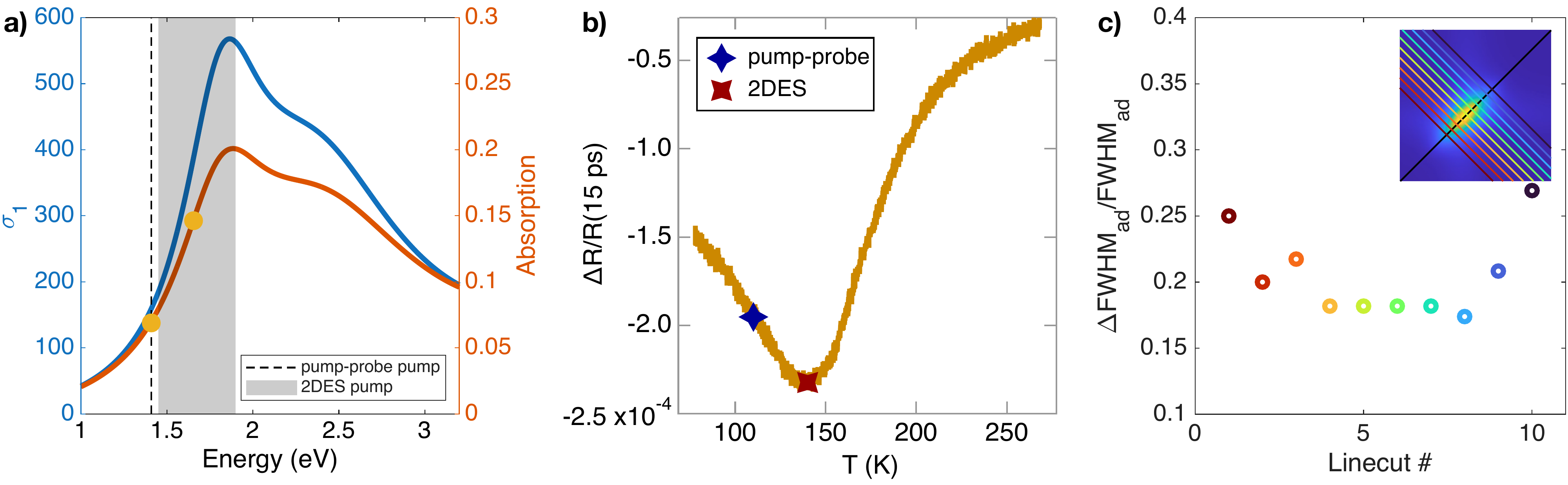}
\caption{a) Low-temperature equilibrium optical conductivity (real part $\sigma_1$) of \ch{LaVO3} (left axis) and corresponding absorption for a 30 nm thin film (right axis). The yellow markers indicate the estimated average absorption at the photon energies used in pump-probe and 2DES. b) Temperature dependence of the transient reflectivity amplitude at probe photon energy 1.77 eV, and at a fixed pump-probe delay $\Delta t=15$ ps (see dynamics in Fig. 1d), highlighting the enhancement of the slow component at temperatures close to the transition. Blue and red markers indicate the temperatures at which pump-probe and 2DES experiments were performed. c) Dependence of the relative anti-diagonal linewidth variation, with respect to equilibrium, on the position of the linecut with respect to peak central position. Inset: simulated 2DES spectrum showing the location of the line cuts taken at different distances from the peak maximum. }
\label{fig: broadening 2}
\end{figure}

The larger relative linewidth change $\Delta\Gamma_{hom}/\Gamma_{hom,0}$ obtained from 2DES ($\sim20\%$) compared to pump-probe measurements ($\sim0.5\%$) is an expected consequence of the different capabilities of the two techniques. Broadband pump-probe spectroscopy probes the total linewidth, whereas 2DES allows the homogeneous component to be disentangled. As a result, substantial pump-induced variations of the homogeneous linewidth can correspond to much smaller relative changes in the total linewidth.
 To verify the compatibility of the two measurements, we simulate a 2DES spectrum using solutions of the optical Bloch equations for a two-level system with Gaussian inhomogeneous broadening \cite{siemens2010resonance}. In the time-domain, the signal is given by
\begin{equation}
\begin{aligned}
s(t_3,t_1) &= s_R(t_3,t_1) + s_{NR}(t_3,t_1) \\
&= S_{0,0} e^{-\Gamma (t_3 + t_1) - i \omega_0 (t_3 - t_1) - \sigma^2 (t_3-t_1)^2/2} \Theta(t_3) \Theta(t_1) \\
    &+ S_{0,0} e^{-\Gamma (t_3 + t_1) - i \omega_0 (t_3 + t_1) - \sigma^2 (t_3+t_1)^2/2} \Theta(t_3) \Theta(t_1)
\label{eq: 2D signal}
\end{aligned}
\end{equation}
where $s_R$ and $s_{NR}$ indicate the repahsing and non-rephasing pulse sequences, respectively.  We consider a resonance at $\omega_0$=1.8 eV, an inhomogeneous width $\sigma$=0.2 eV, and a homogeneous width $\Gamma$=0.09 eV. The resulting 2D spectrum, obtained from the two-dimensional Fourier transform of Eq. \ref{eq: 2D signal}, is shown in Fig. \ref{fig: broadening}a, plotted over a photon-energy range comparable to that of the experiment. We then simulate a pump-induced $20\%$ increase in $\Gamma$ and extract the relative linewidth changes, with respect to equilibrium, in (i) the anti-diagonal line cut of the 2DES spectrum (Fig. \ref{fig: broadening}b) and (ii) the corresponding pump-probe signal, obtained by projecting the 2DES spectrum onto the probe photon energy axis (Fig. \ref{fig: broadening}c). With these parameters, the FWHM of the anti-diagonal line cut increases by $18\%$ compared to equilibrium, whereas the FWHM of the corresponding pump-probe spectrum increases by only $4\%$. \\
On top of this, we need to consider that pump-probe and 2DES measurements were performed under different excitation conditions due to experimental constraints; to properly compare the linewidth variations, the following points must be taken into account.
\begin{itemize}
    \item 
The 2DES experiment used a fluence of 1.4 mJ/cm$^2$ and a broadband pump spectrum spanning 1.45-1.9 eV, whereas the pump-probe measurements were performed with a fluence of 1 mJ/cm$^2$ and a narrowband pump of 1.4 eV central photon energy ($\sim 50$ meV bandwidth). Owing to the stronger absorption of \ch{LaVO3} at 1.8 eV compared to 1.4 eV, these conditions result in markedly different effective excitation densities. Fig. \ref{fig: broadening 2}a shows the parametrized low-temperature equilibrium optical conductivity of \ch{LaVO3} and the corresponding absorption for a 30 nm thin film, indicating that the average absorption at 1.4 eV is approximately half that of the broadband 2DES pump. Accounting also for the lower fluence in pump-probe measurements ($F_\mathrm{pump-probe} \simeq 0.7F_\mathrm{2DES}$), the effective excitation in the pump-probe experiment is estimated to be $\sim$0.35 of that in 2DES.  This reduction is expected to suppress the linewidth variation by at least a factor of three, lowering the estimated pump-probe linewidth change from $4\%$ to $\sim1-1.5\%$.
    \item 
The two experiments were performed at different temperatures (110 K for pump-probe, 140 K for 2DES). At higher temperature, closer to the critical temperature for spin-orbital ordering, the system is more easily perturbed, favoring a larger linewidth change in 2DES. To estimate this effect, we examine the temperature dependence of the transient reflectivity amplitude at a fixed delay ($\Delta t=15$ ps, near the maximum of the slow HE component; see Fig. 1d and Fig. \ref{fig: broadening 2}b). While $\Delta R/R$ includes contributions from both linewidth and spectral-weight changes, the amplitude of the slow component provides a qualitative measure of the enhanced HE linewidth variation at higher temperature, consistent with broadband data and the fits in Fig. 3b. Broadband analysis, available at two selected temperatures (see Fig. 2 and 3), indicates that the $\sim40\%$ reduction in reflectivity amplitude between 110 K and 40 K corresponds to a linewidth variation at 40 K that is $\sim$3 times smaller than at 110~K. Similarly, we can compare 110 K (pump-probe data) to 140 K (2DES data): the smaller $\Delta R/R$ amplitude at 110~K, approximately $\sim20\%$ lower than at 140~K (see markers in Fig. \ref{fig: broadening 2}b), suggests a corresponding reduction in linewidth variation by roughly a factor on the order of 2. This reduces the expected pump-probe linewidth change from $\sim1-1.5\%$ to $\sim0.5-0.7\%$, in good agreement with the measured value.
    \item 
Bandwidth limitations in the 2DES experiments restrict access to only a portion of the HE peak and may prevent the anti-diagonal cut in Fig. 4b from passing exactly through the peak center. Nonetheless, variations in the homogeneous linewidth are expected to affect both central and off-center anti-diagonal line-cuts in a similar manner. To verify this explicitly, the simulated 2DES spectrum of Fig. \ref{fig: broadening} is analysed by extracting anti-diagonal line profiles at different distances from the peak maximum (inset of Fig. \ref{fig: broadening 2}c). The relative linewidth variations with respect to equilibrium are found to be comparable for all cuts, with slightly larger widths variations (up to $\sim40\%$) when the tails of the peak are used for the linecut (see Fig. \ref{fig: broadening 2}c). Therefore, while still providing a reliable representation of relative changes, bandwidth limitations may lead to a slight overestimation of the homogeneous linewidth variation, whose true maximum under the conditions used in the 2DES experiment is expected to lie between $\sim15$ and $20\%$.
\end{itemize}
Overall, these results demonstrate that a substantial difference between linewidth variations extracted from pump-probe and 2DES measurements is expected, and that the experimentally observed variations in 2DES ($\sim20\%$) and pump-probe ($\sim0.5\%$) are quantitatively compatible within this framework.

\clearpage
\section{Supplementary DMFT results}
\begin{figure}[h!]
\centering
\includegraphics[width=7.2cm]{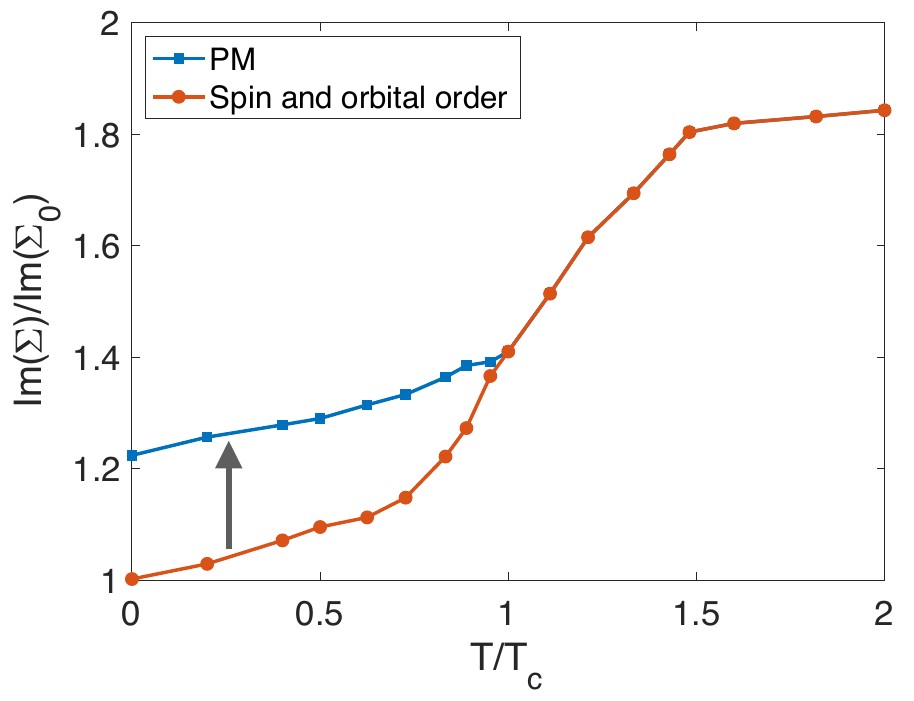}
\caption{
Scattering rate, calculated as the imaginary part of the self-energy $\Sigma$ by means of DMFT, for the cases of the three-band model of \ch{LaVO3} discussed in the main text. The scattering rate grows with temperature, as expected, with a rather clear change occurring at the transition temperature. Suppression of spin and orbital order (blue data points, obtained from the paramagnetic (PM) solution) results in an increase of the scattering rate, highlighted by the gray arrow, as compared to the ordered solution (red data points). The plotted values for $\text{Im}(\Sigma)$ are normalized to the value at $T = 0$~K ($\text{Im}(\Sigma_0)$). 
}
\label{fig: dmft}
\end{figure}

\bibliography{Refs}